\documentclass[letterpaper,twocolumn,10pt]{article}
\usepackage{usenix2019_v3}

\usepackage{amsmath}
\usepackage{graphicx}
\usepackage{booktabs}
\usepackage{enumitem}
\usepackage{url}
\usepackage{xspace}
\usepackage{kotex}
\usepackage[most]{tcolorbox}
\usepackage{pgfplots}
\pgfplotsset{compat=1.18}
\usepackage{color}
\usepackage{subcaption}

\definecolor{color1}{RGB}{70, 130, 180} 
\definecolor{color2}{RGB}{220, 20, 60}  
\definecolor{color3}{RGB}{46, 139, 87}  

\newcommand{\system}{TraceScope\xspace}
\newcommand{\parhead}[1]{\noindent \textbf{#1.}}

\begin{document}

\date{}

\title{\Large \bf TraceScope: Interactive URL Triage via Decoupled Checklist Adjudication}


\author{
Haolin Zhang, William Reber, Yuxuan Zhang, Guofei Gu, and Jeff Huang \\
Texas A\&M University \\
\texttt{\{chris\_zhang, willr, yuz516, jeffhuang\}@tamu.edu}, \texttt{guofei@cse.tamu.edu}
}

\maketitle

\begin{abstract}
Modern phishing campaigns increasingly evade snapshot-based URL classifiers using interaction gates (e.g., checkbox/slider challenges), delayed content rendering, and logo-less credential harvesters. This shifts URL triage from static classification toward an interactive forensics task: an analyst must actively navigate the page while isolating themselves from potential runtime exploits.

We present \system, a decoupled triage pipeline that operationalizes this workflow at scale. To prevent the observer effect and ensure safety, a sandboxed \emph{operator} agent drives a real GUI browser guided by visual motivation to elicit page behavior, freezing the session into an immutable evidence bundle. Separately, an \emph{adjudicator} agent circumvents LLM context limitations by querying evidence on demand to verify a MITRE ATT\&CK checklist, and generates an audit-ready report with extracted indicators of compromise (IOCs) and a final verdict.

Evaluated on 708 reachable URLs from existing dataset (241 verified phishing from PhishTank and 467 benign from Tranco-derived crawling), \system achieves 0.94 precision and 0.78 recall, substantially improving recall over three prior visual/reference-based classifiers while producing reproducible, analyst-grade evidence suitable for review. More importantly, we manually curated a dataset of real-world phishing emails to evaluate our system in a practical setting. Our evaluation reveals that \system demonstrates superior performance in a real-world scenario as well, successfully detecting sophisticated phishing attempts that current state-of-the-art defenses fail to identify.

\end{abstract}


\begin{figure}[t]
  \centering
  \includegraphics[width=\linewidth]{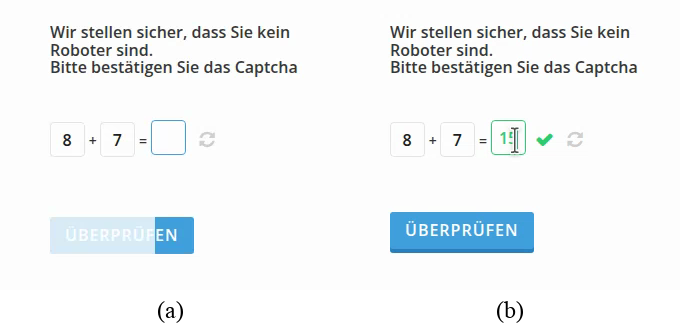}
  \caption{\label{fig:intro-fig} A phishing page protected by a numeric CAPTCHA ($8+7=$). Static classifiers see only the challenge screen (a), which contains no malicious signals. \system visually identifies the arithmetic prompt, computes the correct answer ($15$), and simulates the human keystrokes required to solve the gate (b)}
\end{figure}

\section{Introduction}

Phishing websites remain a dominant initial access vector, yet the web experience that detectors encounter in the wild has fundamentally changed. Beyond simple obfuscation and rapid churn, attackers increasingly deploy interaction gates, legitimate infrastructure abuse, and runtime-only signals to evade detection. These tactics include pages that present only a challenge screen to non-human visitors~\cite{mimecast2024turnstile}, content that appears solely after specific UI interactions, and credential theft implemented as JavaScript-driven form submissions to unrelated endpoints~\cite{sophos2021googleforms, googlecloud2025higheredforms, muncaster2025}. These techniques target a core dependency in the academic URL classification pipeline: the assumption that a model can render a verdict from a static artifact such as a landing-page HTML dump or a single screenshot.

The financial and operational consequences of this sophistication are severe. While Proofpoint reports that telephone-oriented attack delivery (TOAD) threats average roughly 117 million blocked attempts per year~\cite{proofpoint_humanfactor2025}, the monetary impact is escalating through ``wallet drainers'' in the Web3 ecosystem. These attacks, which rely on persuading users to sign malicious transactions, siphoned over \$494 million in 2024 alone~\cite{scamsniffer2024walletdrainers, chainalysis2024drainers}. Similarly, Verizon's 2025 DBIR highlights that human involvement plays a role in 68\% of breaches, driven by a doubling in synthetically generated and socially engineered text~\cite{verizon_dbir2025_infographic}.

Crucially, recent large-scale evaluations highlight that current static collection pipelines are effectively blind to threats due to effective cloaking. Zhang et al. defined this as a ``cat-and-mouse'' game where kits actively profile browser fingerprints to serve benign decoys to security scanners~\cite{zhang2022imspartacus}. This results in a massive ``false downtime'' phenomenon. For instance, Ji et al.~\cite{ji2025_visual_similarity_eval} found that within a corpus of 15.7 million APWG URLs, 61.1\% returned error codes or timeouts to standard crawlers. While this is often attributed to the short lifespan of phishing domains, our deployment data suggests a significant portion represents successful evasion rather than true downtime. Specifically, of the 111 URLs submitted during real-world deployment in our study, 40 were truly offline; among the remaining 71 \textit{active} sites, 18 (25\%) were inaccessible to standard Selenium-based scripts (e.g., returning 403 Forbidden) while remaining fully functional for human users. This implies that the high attrition rates observed in large-scale measurements likely obscure a significant population of active, high-bar threats that cloak themselves against academic scrapers.

This interaction gap is exemplified in Figure~\ref{fig:intro-fig}. In this scenario, a standard crawler or a visual model like Phishpedia~\cite{lin2021phishpedia} would encounter a benign CAPTCHA challenge~\cite{mimecast2024turnstile} or a Vercel-hosted landing page~\cite{cloudflare2026vercel}. Without the ability to perform semantic reasoning and GUI interaction, the detector remains ``blind'' to the malicious payload residing behind the gate. As attackers increasingly adopt these conversational and logic-based gates~\cite{teoh2024phishdecloaker}, existing approaches that rely on lab-environment assumptions fail to see the attack before it vanishes.

Consequently, the practical reality of security operations has shifted. Analysts performing triage must routinely interact with a page to reach a confident verdict, comparing pre- and post-interaction states and tracing exfiltration paths. Our deployment experience reveals three critical shifts in adversary tradecraft that undermine existing automated defenses. First, static HTTP crawlers are routinely fingerprinted and blocked by edge providers. Second, attackers employ targeted cloaking where malicious payloads are only served to visitors matching a specific victim profile. Third, high-volume campaigns increasingly utilize reputation jacking, hosting credential harvesters on legitimate infrastructure (e.g., Google Forms) to bypass domain-reputation filters~\cite{googlecloud2025higheredforms}. This necessitates a transition from lightweight crawling to resource-intensive, full-stack browser emulation that can mimic the victim's environment to reveal the true intent.

However, automating this interactive workflow presents a dilemma. Simply granting an autonomous agent access to the live web introduces severe risks, including observer effects where the act of scanning alters page content, resource exhaustion from infinite navigation loops, and security exposure to drive-by exploits~\cite{provos2008framework}. A robust system must therefore balance the necessity of high-fidelity interaction with the requirements of forensic safety and reproducibility~\cite{nist80086}.

In this work, we present \system, an interactive and trace-driven URL triage system designed to resolve this tension. \system acts as an automated analyst by decomposing triage into two decoupled roles. A persona-driven operator navigates the target using a real GUI browser, guided by visual-semantic goals to elicit hidden behavior and freezing the session into an immutable evidence bundle. Separately, a stateless adjudicator performs structured, checklist-based reasoning over the captured evidence. By separating the high-risk task of interaction from the logic of judgment, \system ensures that forensic verdicts are grounded in immutable evidence, free from the noise and danger of live web execution.

We summarize our contributions as follows:
\begin{enumerate}
    \item We introduce a decoupled forensic architecture that implements a Visual Air-Gap, combining ephemeral sandboxed GUI interaction with strict evidence isolation to analyze evasive pages without exposing the analyst to runtime exploits.
    \item We develop a Deterministic Temporal Normalization layer that solves the challenge of cross-modal synchronization, aligning asynchronous video frames and network packets to prevent hallucinations during evidence retrieval.
    \item We propose a Checklist-Driven Adjudication engine that resolves conflicting evidence via atomic MITRE ATT\&CK verifications, enabling \system to achieve 0.94 precision and 0.78 recall on live URLs while generating audit-ready reports with IOCs.
\end{enumerate}

\section{Background and Related Work}

\paragraph{Traditional Phishing Detection \& Human Factors.}
Early phishing defenses relied heavily on static blocklists and heuristics, a paradigm that remains foundational but increasingly insufficient against evasion.
Kirda and Kruegel introduced \textit{AntiPhish}~\cite{kirda2006antiphish}, a browser extension that tracks sensitive information flow to warn users about untrusted sites, while Provos et al.~\cite{provos2008framework} established early heuristics such as anchor-text mismatches and redirection chains.
To address the scale of attacks, systems like \textit{PhishStorm}~\cite{marchal2014phishstorm} employed streaming analytics on URL features to classify phishing in real-time.
Complementary measurement studies have long highlighted the ecosystem's volatility;
Cao et al.'s \textit{PhishLive}~\cite{cao2012phishlive} and Thomas et al.~\cite{thomas2020gmail} demonstrated that phishing campaigns are often short-lived (measured in hours), necessitating detection speeds that manual blocklisting cannot match.
In the mobile domain, Reaves et al.~\cite{reaves2011sms} and Traynor et al.~\cite{reaves2016legitimate} exposed how UI limitations (e.g., hidden URL bars) and legitimate bulk messaging make SMS phishing particularly effective, a challenge \system addresses by analyzing the network traffic rather than relying on display cues.

A parallel line of research focuses on the ``human factor,'' arguing that technical solutions must account for user behavior.
Herley~\cite{herley2009rational} famously argued that users rationally reject security advice when the cost of compliance outweighs the perceived risk.
Recent work by Sharevski et al.~\cite{sharevski2024accessibility} and Zhao et al.~\cite{tu2019users} reinforces this, showing that users often ignore or misinterpret warnings, whether standard browser alerts or accessibility announcements and remain susceptible to social engineering even when ``trained.''
\system aligns with these insights by offloading the burden of detection from the user to an automated agent that performs the ``victim'' role during triage, generating evidence without requiring human exposure.

\paragraph{AI-Powered and Agentic Approaches.}
As attackers adopted visual obfuscation, defenders shifted toward computer vision and deep learning.
\textit{Phishpedia}~\cite{lin2021phishpedia} and \textit{PhishIntention}~\cite{liu2022phishintention} represent the state-of-the-art in reference-based detection, using deep models to identify brand logos and credential-taking intent.
\textit{DynaPhish}~\cite{liu2023knowledge} extends this by dynamically expanding reference lists to detect brandless phishing.
While effective when content is visible, these systems often struggle against ``interactive cloaking'' where content is hidden behind CAPTCHAs or geofences.
\textit{PhishDecloaker}~\cite{teoh2024phishdecloaker} tackles this specific gap by using AI to solve CAPTCHAs, though it focuses primarily on unmasking rather than comprehensive forensic auditing.
In the email domain, multi-modular systems like \textit{D-Fence}~\cite{lee2021dfence} combine text (BERT) and structural analysis to improve detection, a multi-modal philosophy \system adapts for web analysis.

Recent advances in Generative AI have introduced both new threats and defenses.
Apruzzese et al.'s \textit{E-PhishGen}~\cite{apruzzese2025ephishgen} demonstrates that LLM-generated datasets can expose biases in existing detectors, while other works explore using VLMs for reasoning about page intent and find the brand identifier~\cite{liu2024phishllm}.
However, a gap remains in ``agentic'' defense: active, persona-driven systems that can navigate complex social engineering flows.
While \textit{TRIDENT}~\cite{yang2023trident} uses ML to scan the DOM for social engineering ads, it operates within the user's browser session.
\system builds on general-purpose agent frameworks like Agent~S~\cite{agents} and UI-TARS~\cite{uitars}, but differentiates itself by decoupling interaction from adjudication. By isolating the ``click'' in a sandboxed environment, \system safely elicits hidden malicious behaviors, such as those found in multi-step ``quishing'' or adversarially obfuscated sites, and produces an immutable evidence bundle for audit.


\begin{figure*}[ht!]
  \centering
  \includegraphics[width=\textwidth]{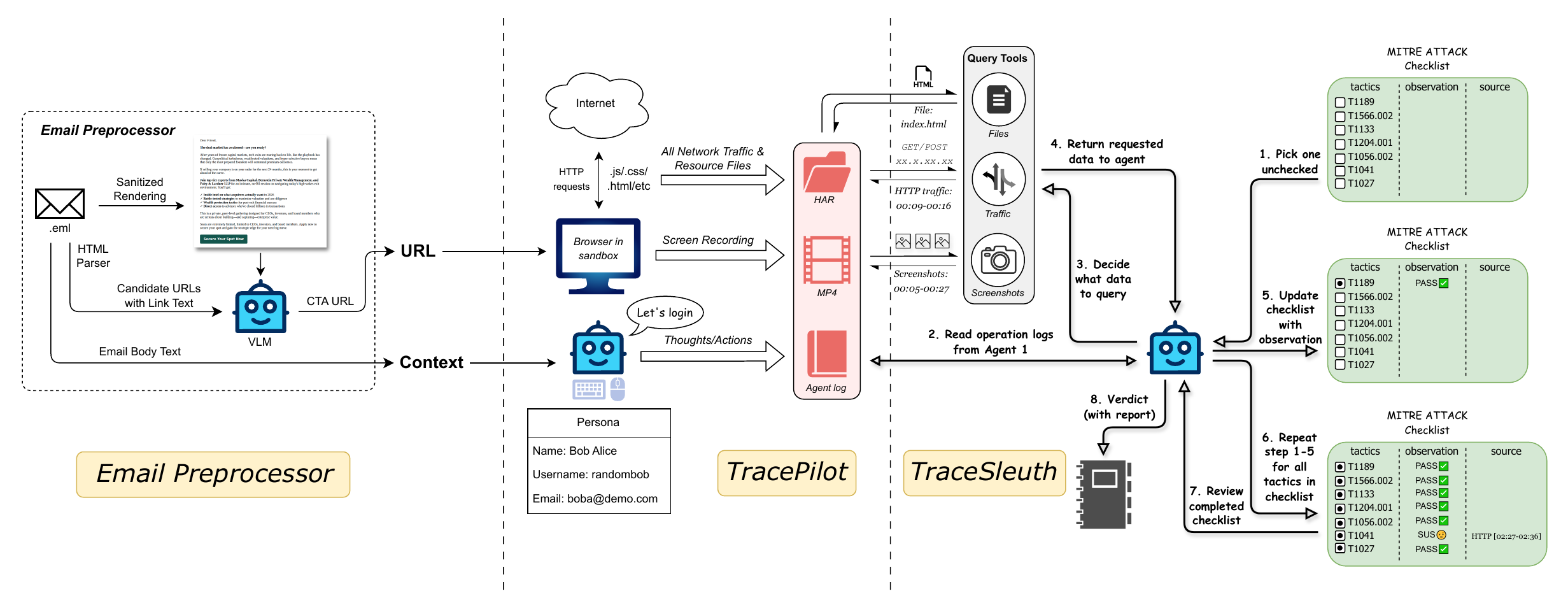}
  \caption{\label{fig:pipeline}\system pipeline. The Operator Agent (TracePilot) interacts with a sandboxed browser to elicit runtime behavior and capture evidence. The Adjudicator Agent (TraceSleuth) queries this evidence on demand to verify a MITRE ATT\&CK checklist, producing a verdict and audit-ready report. Note: While the architecture supports upstream email parsing, this paper evaluates \system starting from URL inputs.}
\end{figure*}

\section{Threat Model and Scope}

We posit a sophisticated web adversary capable of detecting and evading standard automated analysis pipelines. This adversary employs \textit{interactive cloaking} techniques, such as CAPTCHA challenges, behavioral puzzles, and delayed JavaScript rendering to gate malicious content behind checks that block headless crawlers. Furthermore, we consider adversaries who utilize semantic obfuscation to defeat visual classifiers, including the deployment of logo-less credential harvesting forms and the use of conversational interfaces (e.g., support chat simulations) or cryptocurrency wallet interactions that do not trigger traditional credential-requiring page (CRP) detectors. The adversary is assumed to be aware of common scanner IP ranges and user-agent strings, necessitating the use of residential proxies and high-fidelity browser emulation.

\parhead{Defender Goal}
Our primary objective is to replicate the forensic workflow of a human analyst in a secure, automated environment. Given a candidate URL and its lure context, the system must (i) actively navigate through interaction gates to elicit the underlying page intent, (ii) resolve the final verdict (Phishing/Benign) based on runtime evidence rather than static snapshots, and (iii) synthesize an evidence-backed audit report. This report must contain verifiable Indicators of Compromise (IOCs) mapped to the MITRE ATT\&CK framework, suitable for immediate consumption by IR teams for takedowns or blocklisting updates.

\parhead{Attacker's Capability}
We assume the adversary controls the full serving stack and can execute arbitrary server-side logic to determine whether to serve a malicious payload or a benign decoy. Specifically, the attacker is capable of: (i) \textit{Client-side Fingerprinting}, utilizing JavaScript to inspect browser environmental variables, hardware concurrency, and even mouse movement trajectories to identify automated sandboxes; (ii) \textit{Network-level Filtering}, performing geofencing or blocklisting based on known data center IP ranges to isolate victims from security vendors; and (iii) \textit{Interaction-Gating}, requiring specific UI state transitions, such as solving a numeric CAPTCHA or engaging in a multi-turn chat before injecting the credential-harvesting DOM elements. We further assume the attacker can monitor exfiltration endpoints for repeated use of test credentials, allowing them to burn known research accounts and evade detection by maintaining a high ``signal-to-noise'' ratio of real victims.

\parhead{Safety and Boundary Conditions}
We enforce a strict safety posture where the operator agent utilizes synthetic personas with valid formats but no real-world value, effectively insulating the organization from credential leakage. All execution occurs within ephemeral sandboxed browser environments that are discarded post-session to mitigate drive-by compromise risks. While TraceScope targets interactive evasion, we consider hard cryptographic challenges (e.g., MFA token requirements) and region-locked content requiring specific physical geolocations out of scope. 

\parhead{Deployment Scenario: Cascading Defense}
To address the operational trade-off between inspection depth and latency, we define TraceScope's role within a layered security architecture. It is not designed as a high-throughput perimeter filter for synchronous blocking of all web traffic. Instead, TraceScope operates as an on-demand forensic agent triggered by upstream signals, such as user-reported phishing emails or ``suspicious/uncertain'' verdicts from lightweight static classifiers. In this context, the system serves to automate the labor-intensive Triage and Incident Response (IR) phase. While its computational cost is higher than static analysis, it provides orders-of-magnitude reduction in cost and latency compared to the human analyst it replaces, who would otherwise manually investigate the URL, navigate evasion gates, and compile an incident report.

\section{System Design}

\subsection{Overview}
Figure~\ref{fig:pipeline} illustrates the \system architecture, which is designed to reconcile the conflicting requirements of high-fidelity interaction and forensic safety. To achieve this, the system decouples the analysis lifecycle into three isolated stages: safe intention resolution, sandboxed emulation, and post-hoc adjudication. This separation of concerns ensures that forensic analysis does not influence the runtime behavior of the target, effectively eliminating the observer effect common in live introspection.

The pipeline begins by addressing the risk of pre-click exploitation. The Email Preprocessor acts as a visual air-gap, ingesting raw \texttt{.eml} files to determine the analysis entry point without executing potential payloads. Rather than relying on fragile regex parsers, the module performs a \textit{sanitized rendering} of the email body, neutralizing active scripts and trackers, and employs a Vision-Language Model (VLM) to semantically identify the intended Call-to-Action (CTA). This ensures the system captures the ``victim's view'' of the threat, distinguishing prominent malicious buttons from benign footer links before any network connection is made.
Once a target is identified, TracePilot (Operator) executes the interaction. Unlike traditional scanners that attempt to analyze content in real-time, TracePilot is purely operational. It assumes a specific user persona to navigate the target site, handling deceptive flows such as CAPTCHAs or multi-step redirections within a disposable sandbox. Its primary objective is not to judge, but to generate an immutable ledger of the session, capturing full-packet network traffic (HAR) and visual evidence (video/screenshots), thereby freezing the attack state for safe downstream consumption.
The final verdict is derived by TraceSleuth (the Adjudicator), a forensic agent that operates exclusively on the recorded artifacts. By isolating the adjudication logic from the browsing environment, \system ensures reproducibility; the same session evidence can be re-evaluated against updated threat intelligence without re-visiting the (potentially transient) malicious infrastructure. TraceSleuth systematically verifies the session against a MITRE ATT\&CK-mapped checklist, querying time-sliced network logs and visual feeds to produce a comprehensive, audit-ready report.

\subsection{Email Preprocessor: Target Extraction}
\label{sec:preprocessor}

The initial phase of the pipeline addresses the challenge of \textit{Intention Resolution} by identifying the adversary's intended victim path while insulating the analysis infrastructure from active threats. This module operates as a visual air-gap that utilizes a visual-semantic selection strategy to isolate the primary Call-to-Action (CTA).

To mitigate the risks associated with analyzing malicious emails, including drive-by downloads or tracking pixels, the system converts the email body into an inert visual representation. The raw HTML content is parsed and neutralized to prevent outbound network requests while preserving the structural integrity of the layout. This safe DOM is rendered within a headless browser to capture a high-fidelity screenshot. This process ensures the system acquires the visual features necessary for analysis, such as button styling and layout hierarchy, without executing potential exploits.

Concurrently, the preprocessor extracts a list of candidate hyperlinks from the DOM. Traditional phishing detection frequently fails because regex-based parsers cannot distinguish between functional links, such as unsubscribe or privacy policy pages, and the malicious CTA, which is often masked by generic text. To resolve this ambiguity, \system leverages a Vision-Language Model (VLM). The model receives the inert screenshot alongside the enumerated list of links, as depicted in Figure~\ref{fig:vlm_prompt}. By prompting the VLM to identify the primary Call-to-Action, the system exploits the capacity of the model to correlate visual prominence, such as a large or centered button, with a specific URL index. This visual-semantic alignment enables the preprocessor to pinpoint the intended victim path even when the link text is obfuscated.
Upon isolation of the target URL, the preprocessor extracts the plain text body of the email to construct the \textit{Session Context}. This context is critical for the subsequent emulation phase. It is forwarded to the TracePilot agent to provide the semantic framing, such as an urgent invoice payment, required to simulate a motivated victim rather than a nondescript crawler.

\subsection{TracePilot: High-Fidelity User Emulation}
\label{sec:tracepilot}

\begin{figure}[ht!]
\centering

\begin{tcolorbox}[
    width=0.92\linewidth,
    colback=blue!5,
    colframe=blue!60,
    title=\textbf{System Prompt: TracePilot Emulation},
    fonttitle=\bfseries\sffamily,
    boxrule=0.4mm,
    arc=1.5mm,
    boxsep=1mm,
    left=1mm,
    right=1mm,
    top=1mm,
    bottom=1mm,
]

\footnotesize\ttfamily
\linespread{0.94}\selectfont

You reached this website from the message below. Behave like a normal user following the message's instructions and embedded links.\\[0.3em]

\textbf{Message context:}\\
---\\
Subject: Action Required: Update your billing information\\
From: support@streaming-service.com\\
Body: Dear Customer, we could not process your latest payment...\\
---\\[0.3em]

\textbf{Persona (use only these details; do not invent data):}\\
- Full name: Bob Alice\\
- Username: randombob\\
- Password: ZK29YcCITMb!\\
- Email: boba@demo.com\\
- Phone: +1-555-0199\\
- Address: 123 Maple Ave, Springfield, IL 62704, US\\
- Payment card: Visa 1234 5678 9012 3456 exp 12/29 CVC 000\\
- Browser UA: Mozilla/5.0 (Macintosh; Intel Mac OS X 10\_15\_7)...\\[0.3em]

\textbf{Behavior rules:}\\
- Act like an ordinary user: click obvious calls-to-action, scroll, read...\\
- If asked to sign in, use the persona's username/password above.\\
- If a form asks for payment or identity info, fill fields with persona data.\\
- Never use data not listed in the persona block; never upload files.\\
- When a website asks to open an application, do not allow it.

\end{tcolorbox}

\caption{Structured system prompt used to initialize TracePilot, combining email context with a synthetic persona and behavioral guardrails.}
\label{fig:tracepilot_prompt}

\end{figure}


To penetrate evasion techniques that fingerprint and block automated scanners, the pipeline transitions to TracePilot. This module is an LLM-driven agent responsible for high-fidelity user emulation. Unlike traditional crawlers that mechanically traverse the DOM, TracePilot adopts a distinct ``victim persona,'' operating with specific motivations and behavioral patterns defined by the system instruction in Figure~\ref{fig:tracepilot_prompt}.


\parhead{Evolving Phishing Paradigms}
The landscape of phishing has evolved far beyond traditional credential-harvesting pages, rendering existing static and snapshot-based defenses increasingly obsolete. Modern adversaries have pioneered new attack vectors, such as the rise of sophisticated Web3 wallet drainers that specifically target digital assets through malicious smart contract interactions, resulting in nearly half a billion dollars in losses~\cite{scamsniffer2024walletdrainers, chainalysis2024drainers}. Furthermore, even traditional phishing has adopted advanced cloaking and verification techniques, such as Cloudflare Turnstile or customized CAPTCHAs, to hide malicious payloads from automated crawlers~\cite{mimecast2024turnstile, zhang2022imspartacus}. Attackers also increasingly abuse legitimate platforms, such as Google Forms, to host malicious lures on trusted domains that bypass reputation-based filters~\cite{sophos2021googleforms, muncaster2025}. These trends create a fundamental challenge for existing security measures: the malicious intent is often hidden behind multiple layers of benign-looking interaction. Consequently, there is an urgent need for a defensive framework capable of active, human-like exploration to successfully peel back these layers and reveal the underlying harmful intentions of a suspicious website.

LLM agents emerge as a uniquely powerful solution for navigating these interactive phishing paradigms due to their advanced multimodal and reasoning capabilities. Unlike heuristic-based bots, LLM agents possess a holistic understanding of both visual layout and semantic text, allowing them to interpret complex UI components and social engineering lures with human-level nuance~\cite{yang2023dawn}. This multimodal intelligence is augmented by sophisticated reasoning chains, which enable agents to solve the various verification challenges that phishers use to cloak their content from defenders, such as puzzle-based CAPTCHAs or dynamic redirects~\cite{shinn2023reflexion}. Moreover, by leveraging tool access to drive browser environments directly, these agents can autonomously interact with page elements to trigger the final malicious payload that typical crawlers miss~\cite{zhou2023webarena}. Recent research underscores that agentic workflows significantly outperform traditional classifiers in high-entropy environments by decoupling high-level intent from low-level execution~\cite{weng2023llm}. By utilizing these capabilities, \textsc{TracePilot} can adapt to novel attack variants in real-time, providing the forensic depth necessary for robust URL triage.

\parhead{Sandboxed Execution and Anti-Fingerprinting}
A critical operational challenge in modern phishing analysis is the fragility of long-running automation and the aggressive blocking of headless browsers. Previous studies indicate that up to 61.1\% of phishing URLs sourced from feeds like APWG are unreachable or fail to resolve when accessed by standard crawlers~\cite{ji2025_visual_similarity_eval}. Our own deployment experience corroborates this; we observed that 18 out of 71 live URLs returned HTTP 403 Forbidden or 500 Service Error responses when accessed via a standard headless Selenium driver, yet opened successfully in a consumer-grade browser.

To address these availability gaps, TracePilot moves beyond simple headless emulation. It operates within a containerized \texttt{linuxserver/webtop}\cite{linuxserver_webtop} environment that renders a full X11 desktop session. To prevent the resource exhaustion and zombie processes common in long-running crawls, we implement a ``disposable browser'' architecture. The orchestration server spawns a dedicated subprocess for each URL analysis task, enforced by a strict OS-level watchdog. If a session exceeds its allocated time window or hangs due to adversarial scripts, the watchdog terminates the entire process group (killpg). This ensures that every analysis begins with a pristine, unpolluted memory state and network stack, significantly reducing the surface area for fingerprinting.

\parhead{Immutable Evidence Capture}
Standardizing evidence across the network, visual, and cognitive layers is difficult because these data streams run asynchronously. A network request may occur milliseconds before the corresponding DOM update, making it difficult to causally link an agent's click to a specific exfiltration packet.

TraceScope solves this by decoupling the recording logic from the agent's decision loop. While the agent navigates, a dedicated background service captures the session into an immutable bundle:
\begin{itemize}
    \item \textbf{Visual Layer:} Instead of relying on DOM snapshots which miss native popups, we employ an \texttt{ffmpeg} process hooked directly to the X11 display buffer. This produces a high-fidelity video recording that captures the exact visual experience of the victim, including cloaking animations and external redirects.
    \item \textbf{Network Layer:} A synchronized HAR (HTTP Archive) recorder captures full-packet traffic, preserving headers and payloads for downstream forensic analysis.
    \item \textbf{Cognitive Trace:} To explain \textit{why} an action occurred, we implement a ``PyAutoCapture'' shim that intercepts low-level inputs. This module injects visual overlays (e.g., coordinate markers for clicks, bounding boxes for typing) onto the evidence screenshots and generates a markdown-formatted audit log.
\end{itemize}
This multi-modal approach eliminates the observer effect and ensures that downstream forensic agents analyze a pristine record of the attack as it occurred, unaffected by the latency or overhead of real-time inspection tools.

Figure~\ref{fig:tracepilot_prompt} illustrates the system prompt for TracePilot, which is engineered to transform a general-purpose LLM into a targeted emulation agent through four functional blocks. First, the Contextual Anchor aligns the agent's intent with the specific phishing lure (e.g., billing updates). Second, a Synthetic Persona provides disposable PII to satisfy data-harvesting forms without compromising real user data. Third, Behavioral Directives enforce ordinary user patterns, such as scrolling and clicking, to bypass bot detection. Finally, Operational Guardrails prevent high-risk actions like file uploads or application launches, ensuring the agent remains safely contained within the sandbox during analysis.

\subsection{TraceSleuth: Forensic Adjudication}
\label{sec:tracesleuth}

\begin{table*}[ht!]
\centering
\begin{tabular*}{\textwidth}{@{\extracolsep{\fill}}lcll@{}}
\toprule
\textbf{Profile} & \textbf{\# Tech.} & \textbf{ID} & \textbf{Description} \\
\midrule
Minimal & 7 
  & T1189 & Drive-by Compromise \\
& & T1566.002 & Phishing Link \\
& & T1133 & External Remote Services \\
& & T1204.001 & Malicious Link Execution\\
& & T1056.002 & Input Capture: GUI Input Capture \\
& & T1041 & Exfiltration Over C2 Channel \\
& & T1027 & Obfuscated Files or Information \\
\midrule
Standard & 12 
  & \textit{--} & \textit{Includes all Minimal techniques, plus:} \\
& & T1059.007 & Command and Scripting Interpreter: JavaScript \\
& & T1071.001 & Application Layer Protocol: Web Protocols \\
& & T1102 & Web Service \\
& & T1078 & Valid Accounts \\
& & T1098 & Account Manipulation \\
\midrule
Comprehensive & 14 
  & \textit{--} & \textit{Includes all Standard techniques, plus:} \\
& & T1110.003 & Brute Force: Password Spraying \\
& & T1539 & Steal Web Session Cookies \\
\bottomrule
\end{tabular*}
\caption{\label{tab:checklist_profiles}Checklist profiles used by the adjudicator. The table details the specific MITRE ATT\&CK techniques included in each profile. \textbf{Minimal} focuses on high-signal phishing behaviors. \textbf{Standard} adds credential harvesting and delivery tradecraft. \textbf{Comprehensive} encompasses the full playbook, including noisier techniques like ARP poisoning.}
\end{table*}


\begin{figure}[ht!]
\centering
\begin{tcolorbox}[
width=0.98\linewidth,
colback=red!5,
colframe=red!60,
title=\textbf{System Prompt: TraceSleuth Forensics},
fonttitle=\bfseries\sffamily,
boxrule=0.35mm,
arc=1.5mm,
boxsep=0.8mm,
left=0.8mm, right=0.8mm,
top=0.6mm, bottom=0.6mm,
titlerule=0pt,
before skip=0pt,
after skip=2pt
]
\footnotesize\ttfamily
\linespread{0.92}\selectfont

You are \textbf{TraceSleuth}, a security investigation specialist focused on high-signal phishing detections from captured browser sessions.\\[-0.2em]

\textbf{Protocol:}~
Guided through one MITRE ATT\&CK technique at a time;
base conclusions \textbf{strictly} on artifacts pulled through tools;
never speculate without a citation to a specific artifact;
clearly map observations back to the requested technique.\\[-0.2em]

\textbf{Available MCP Tools:}~
\texttt{get\_session(time, filter)} (HTTP traffic/headers);
\texttt{get\_screenshot(time)} (rendered frames);
\texttt{retrieve\_resource(prefix)} (persisted HTML/JS artifacts).\\[-0.2em]

\textbf{Output Format (JSON):}\\[-0.2em]
\{\\
\ \ ``status'': ``confirmed'' | ``suspicious'' | ``not\_observed'',\\
\ \ ``confidence'': ``high'' | ``medium'' | ``low'',\\
\ \ ``evidence'': [ \{ ``source'': ``...'', ``observation'': ``...'', ``relevance'': ``...'' \} ]\\
\}

\end{tcolorbox}

\caption{The forensic instruction set for TraceSleuth. The agent is constrained to act as an objective auditor, accessing data solely through the provided MCP tool definitions and outputting structured verdicts mapped to MITRE ATT\&CK techniques.}
\label{fig:tracesleuth_prompt}
\end{figure}

While TracePilot successfully captures the raw artifacts required to reveal a URL's underlying intent, existing detection frameworks often fail to interpret these signals correctly. The primary failure mode stems from a fundamental shift in adversarial strategy: to evade popular logo-based and reference-matching detectors, modern attackers increasingly deploy logo-less phishing sites that purposefully omit brand-specific visual cues, thereby failing to trigger standard similarity-based alerts. Furthermore, as previously discussed, emerging paradigms like crypto-wallet drainers bypass the traditional credential-request heuristic by soliciting direct wallet authorizations or smart contract interactions rather than plaintext passwords. These scenarios render defenses that rely on the dual assumptions that phishing must (a) mimic a known enterprise and (b) trivially harvest credentials, increasingly ineffective. To address these sophisticated evasions, we propose TraceSleuth, which mimics the investigative process of a human security expert. By utilizing an LLM agent as a forensic analyst, we provide the necessary context and a multidimensional security feature checklist to guide its reasoning. This approach moves beyond simple pattern matching, allowing the agent to synthesize diverse evidence from DOM anomalies to behavioral triggers into a high-fidelity security verdict that remains robust against non-traditional and logo-less attack vectors.


\parhead{The Stateless Auditor}
The agent executes a structured forensic playbook aligned with the MITRE ATT\&CK framework~\cite{mitre_attack}. To mitigate the hallucination risks inherent in open-ended LLM queries, TraceSleuth operates strictly as a ``Stateless Auditor'' via the Model Context Protocol (MCP). It processes techniques individually, such as T1189 (Drive-by Compromise) or T1566.002 (Phishing Link), using a read-only toolset that allows it to pull specific data slices on demand. This design enforces \textit{tactic-evidence coupling}, a mechanism that prevents the model from speculating on a verdict without citing a specific resource ID or video timestamp from the TracePilot session. To accommodate diverse operational requirements, the adjudication logic is segmented into scalable profiles. These range from ``Minimal'' settings that focus on high-signal phishing to ``Comprehensive'' configurations that iterate through complex heuristic checks defined in the system's JSON definitions, allowing operators to balance token costs against the required level of forensic depth.

\parhead{Cross-Modal Synchronization via TraceView}
A critical challenge in multi-modal forensics involves aligning continuous video streams with discrete network events without relying on the unreliable internal clock of an LLM. While TracePilot captures the raw asynchronous streams, TraceSleuth employs a dedicated middleware, TraceView, to perform deterministic temporal normalization. Upon loading a session, the server establishes a global epoch ($T_0$) based on the initial network handshake. It then abstracts raw timestamps into relative epochs (e.g., $T+15.2s$). When the adjudicator queries a specific interaction event, the API performs a dual-seek operation: it retrieves the mathematically aligned visual state (Frame $F_t$) via accurate seeking and filters the network log for a concurrent burst (Requests $R_{t \pm 0.5s}$). This strict $\pm 0.5s$ temporal window ensures the model reasons over temporally locked artifacts, preventing ``temporal hallucinations'' where the agent conflates a login popup with an unrelated background request.

\parhead{Resolving Conflicting Evidence}
Live phishing sites often present contradictory signals, such as a visually authentic Microsoft login page hosted on a legitimate but compromised WordPress domain. Monolithic classifiers frequently fail in these scenarios because they are forced to choose between the safe domain reputation and the malicious visual content. TraceSleuth resolves this ambiguity via Checklist-Driven Atomicity. The Adjudicator does not render a single binary verdict; instead, it iterates through isolated MITRE techniques, checking T1041 (Exfiltration Over C2 Channel) separately from T1078 (Valid Accounts). The system enforces a strict Evidence Citation Protocol, under which the model cannot speculate on a verdict unless it populates a structured evidence object with a specific resource ID. This requirement transforms conflicting evidence from a failure mode into a granular forensic finding that captures the divergence as a verifiable feature of the attack.

\parhead{Synthesis \& Reporting}
While JSON verdicts suffice for automated blocking rules, human analysts and security executives require semantic context to make informed decisions about containment and remediation. Raw technical logs often lack the narrative cohesion necessary to assess the broader risk, such as the potential blast radius or the sophistication of the adversary. To bridge this gap, the final phase of the pipeline activates the Report Writer, an LLM-driven agent initialized with the specific constraints defined in the system prompt. This component serves as a semantic translation layer that synthesizes the fragmented technical findings, such as isolated MITRE technique detections and raw network timestamps into a polished Incident Response document. Unlike standard summarization tools that could hallucinate details or repeat facts, the Report Writer enforces a strict ``Bottom Line Up Front'' structure. It prioritizes a high-confidence Executive Summary and a ``Security To-Do List'' containing specific remediation steps, such as blocking specific domains or resetting compromised user credentials. Meanwhile, the system relegates granular technical proofs, including the Evidence Cross-Reference and full MITRE ATT\&CK Mapping, to the appendices. This deliberate organization ensures the final output constitutes actionable intelligence that allows security operations teams to immediately understand the ``who, what, and where'' of the attack without wading through pages of raw JSON data.

The system prompt for TraceSleuth is detailed in Figure~\ref{fig:tracesleuth_prompt}. The prompt establishes a strictly evidentiary protocol, requiring the agent to adjudicate threats by investigating one MITRE ATT\&CK technique at a time. To ensure factual grounding and eliminate speculative reasoning, the agent is restricted to an immutable evidence bundle accessed via specific Model Context Protocol (MCP) tools for retrieving HTTP traffic, screenshots, and source code artifacts. Finally, the prompt mandates a structured JSON output, forcing the agent to categorize its findings by status and confidence level while providing direct citations for every observation. This architecture ensures that the final verdict is not merely a classification, but a reproducible forensic report that maps observed runtime behaviors to established adversarial tactics.

\section{Evaluation}

We organize evaluation around the following questions, inspired by the structure used in prior systems such as PhishLLM~\cite{liu2024phishllm}.
\begin{itemize}[leftmargin=*,nosep]
\item \textbf{RQ1:} How does TraceScope perform relative to prior phishing URL classifiers on \emph{live} URLs?
\item \textbf{RQ2:} Which failure modes cause snapshot classifiers to miss phishing in deployment (e.g., interaction gates, logo-less harvesters), and how does TraceScope address them?
\item \textbf{RQ3:} What is the operational efficiency and economic viability of interactive triage?
\item \textbf{RQ4:} How effective is \system when deployed in a real-world production environment against analyst-verified threats?
\end{itemize}

\begin{table}[htpb]
\centering
\setlength{\tabcolsep}{3pt}
\renewcommand{\arraystretch}{1.05}
\begin{tabular}{@{}p{0.44\columnwidth} r p{0.42\columnwidth}@{}}
\toprule
\textbf{Stage} & \textbf{URLs} & \textbf{Notes} \\
\midrule
Scheduled \& reachable & 1000 & DNS-resolvable and non-404 (531 phishing / 469 benign). \\
Excluded: label drift & 236 & URL reachable, but content removed/suspended/replaced. \\
Excluded & 56 & Static snapshot captured late-render/loading state. \\
\midrule
\textbf{Evaluated (Final Set)} & \textbf{708} & \textbf{241 phishing / 467 benign.} \\
\bottomrule
\end{tabular}
\caption{\label{tab:funnel}Evaluation funnel and symmetric exclusions applied to all methods. Excluded mean no meaningful content and therefore removed from the dataset.}
\end{table}

\subsection{Experimental Setup}
\label{sec:setup}

\begin{figure*}[ht!]
  \centering
  \includegraphics[width=\textwidth]{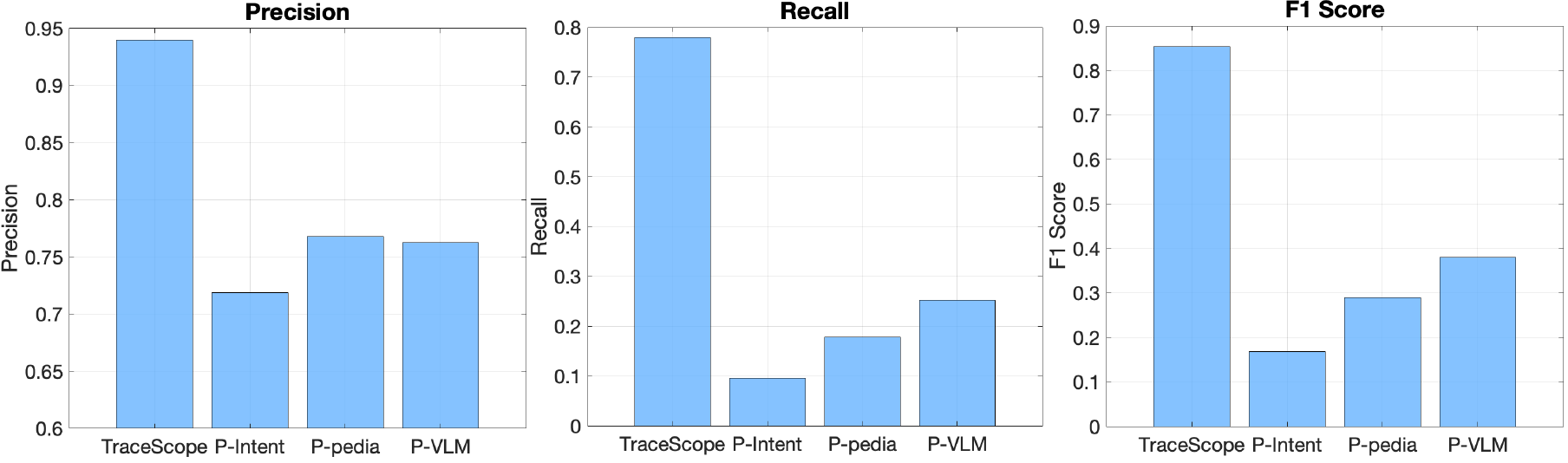}
  \caption{\label{fig:rq1} Detection performance comparison between \system and state-of-the-art baselines on live urls (P-Intent means PhishIntention\cite{liu2022phishintention}, P-pedia means Phishpedia\cite{lin2021phishpedia}, P-VLM menas PhishVLM\cite{liu2024phishllm}). \system shows higher precision while maintaining a significant higher recall. This indicate \system introduces lower false negatives and showcases stronger capability in capturing phishing urls compared to existing baselines. }
\end{figure*}

\parhead{Datasets and Baselines} 
We evaluate \system on live URLs drawn from two sources: phishing URLs from PhishTank~\cite{phishtank} (restricted to ``online, verified'' entries) and benign URLs derived from the Tranco top-sites list~\cite{tranco}. To ensure a realistic evaluation, we crawl within Tranco domains to obtain long-tail sub-URLs rather than testing only root domains. Before evaluation, a reachability filter removes dead URLs (e.g., DNS failures or HTTP 4xx/5xx errors).

We compare \system against three state-of-the-art phishing classifiers: PhishIntention~\cite{liu2022phishintention}, Phishpedia~\cite{lin2021phishpedia}, and PhishVLM~\cite{liu2024phishllm}. All baselines receive the same static inputs: the URL string, the landing-page HTML, and a rendered screenshot from a headless Selenium sandbox. These baselines run inside an isolated Docker environment following their official repositories, while \system operates on the host system.

\parhead{Live Evaluation Pipeline}
To minimize churn bias and ensure paired comparisons, we utilize a continuous live-URL pipeline. An automated coordinator samples a previously unseen URL every 5 minutes from a near-balanced pool of phishing and benign candidates. File locks ensure that all four systems score the \emph{same} live instance of a URL before the next sample is issued.

\parhead{Evaluation Funnel and Churn Handling}
Phishing feeds are highly volatile; many URLs are taken down minutes after report. To maintain a fair comparison, we apply two \emph{symmetric exclusions} that prevent any system from observing the intended content (Table~\ref{tab:funnel}). 
First, we exclude 235 URLs that were reachable but no longer showed the reported phishing content due to suspension, takedown, or indirection via shorteners (\emph{label drift}). Second, we exclude 56 URLs where the headless snapshot captured only an intermediate loading state or animation (timing/late-render issues), as this would unfairly disadvantage static baselines. 
From an initial 1,000 scheduled runs, these exclusions yield a final evaluated set of \textbf{708} reachable-and-analyzable URLs (241 phishing, 467 benign).

\parhead{Metrics}
We report precision, recall, and F1-score for binary phishing detection. Errors (e.g., model endpoint failures) are tracked separately and excluded from the confusion matrix. Because precision is sensitive to the underlying class prior, we also model performance across varying benign-to-phishing ratios to simulate different deployment environments.

\subsection{RQ1: Detection Performance on Live URLs}
\label{sec:rq1}

Figure~\ref{fig:rq1} illustrates the overall performance across the 708 evaluated URLs (241 phishing, 467 benign). \system achieves an F1-score of 0.8507, significantly outperforming all snapshot-based baselines. As the results indicate, the most substantial gain is observed in recall (0.7769 vs. 0.2521 for PhishVLM), confirming that interactive evidence collection successfully exposes phishing intent that remains hidden in static snapshots. While baselines maintain moderate precision in specific contexts, their failure to bypass interaction gates or resolve late-loading content results in prohibitive false-negative rates.

Our evaluation set is phishing-enriched ($\approx$34\% prevalence), mimicking a specialized SOC triage queue. However, since real-world deployment environments vary, ranging from perimeter filters with rare phishing to abuse mailboxes with high prevalence, we model the F1-score stability across varying traffic ratios in Figure~\ref{fig:prior_shift}. We hold each method's true-positive and false-positive rates fixed while varying the assumed phishing prevalence $\pi$. As shown in the performance curves, \system maintains superior stability and a higher F1-score across all prevalence ratios compared to PhishIntention, Phishpedia, and PhishVLM. This profile confirms that \system is optimized as an analyst-grade triage system for enriched queues rather than a general-purpose first-pass filter.

\subsection{RQ2: Analysis of Failure Modes}
\label{sec:rq2}

To understand \emph{why} snapshot-based baselines fail where \system succeeds, we performed a post-hoc manual categorization of the evaluable phishing URLs. A human analyst labeled each phishing URL into one of five structural categories (Table~\ref{tab:buckets}).

\begin{table*}[t]
\centering
\setlength{\tabcolsep}{0pt}
\begin{tabular*}{\textwidth}{@{\extracolsep{\fill}} l r c c c c @{}}
\toprule
& & \multicolumn{4}{c}{\textbf{Recall by Method}} \\
\cmidrule(l){3-6}
\textbf{Phishing Category} & \textbf{Count} & \textbf{PhishIntention} & \textbf{Phishpedia} & \textbf{PhishVLM} & \textbf{\system} \\
\midrule
\textbf{Brand Impersonation} & 76 & 5.3\% & 30.3\% & 31.6\% & \textbf{73.7\%} \\
\textit{(Login/Domain mismatching)} & & & & & \\
\midrule
\textbf{Logo-less / Generic} & 16 & 0.0\% & 0.0\% & 12.5\% & \textbf{56.3\%} \\
\textit{(No/Weak brand assets; text-only lures)} & & & & & \\
\midrule
\textbf{Non-CRP Capture} & 12 & 0.0\% & 8.3\% & 8.3\% & \textbf{100.0\%} \\
\textit{(Crypto wallet; PII)} & & & & & \\
\midrule
\textbf{Interaction Gated} & 12 & 0.0\% & 16.7\% & 8.3\% & \textbf{16.7\%} \\
\textit{(CAPTCHAs; interstitial blocks)} & & & & & \\
\midrule
\textbf{Progressive Disclosure} & 2 & 0.0\% & 0.0\% & 0.0\% & \textbf{50.0\%} \\
\textit{(Intent hidden behind chat/clicks)} & & & & & \\
\bottomrule
\end{tabular*}
\caption{\label{tab:buckets}Recall breakdown by phishing tactic. \system dominates in ``Logo-less'' and ``Non-CRP'' scenarios where visual baselines, including PhishIntention~\cite{liu2022phishintention}, Phishpedia~\cite{lin2021phishpedia}, and PhishVLM~\cite{liu2024phishllm} lack signal. Counts reflect the subset of evaluated URLs manually labeled for this analysis.}
\end{table*}

Table~\ref{tab:buckets} reveals that while \system improves performance on standard brand impersonation (73.7\% vs. 31.6\% for the best baseline), the performance gap is most dramatic in categories where visual branding is absent or the objective is not a standard credential pair (Non-CRP). PhishIntention performs particularly poorly on this live set (5.3\% on brand impersonation), likely due to its strict OCR-and-layout matching requirements which fail when live phishing kits employ slight obfuscations or modern CSS rendering tricks.

\subsubsection{Addressing Logo-less and Generic Phishing}
Visual classifiers like Phishpedia and PhishVLM rely on detecting high-confidence logo bounding boxes. Attackers evade this by using generic ``security update'' templates that omit trademarked logos. In our breakdown, baselines achieved $\approx$0--12\% recall on this category, while \system achieved 56\%.

\begin{figure}[t]
  \centering
  \includegraphics[width=\linewidth]{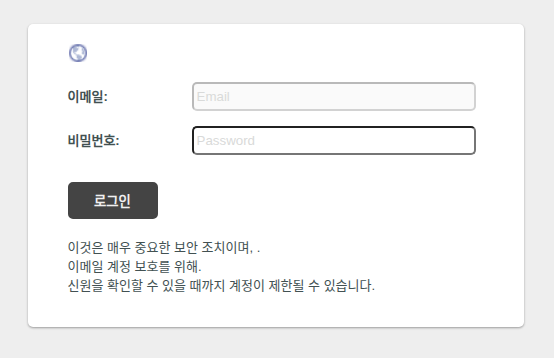}
  \caption{\label{fig:case-logoless}Case study: An example of logoless credential harvest phishing targeting Korean speakers. The attack avoids branding to bypass visual filters. The fields are labeled 이메일 [Email] and 비밀번호 [Password]. The text at the bottom uses faux urgency: ``This is a very important security measure... Your account may be restricted until your identity can be verified.''}
\end{figure}

\parhead{Case Study: Logo-less Credential Harvesters}
Figure~\ref{fig:case-logoless} illustrates a representative failure mode targeting Korean speakers. The page uses a minimal design with fields labeled \emph{이메일} (Email) and \emph{비밀번호} (Password) and urgency text (``Account restricted''), but displays no organization name or logo.
Static baselines under-trigger because the visual signal is too generic. \system correctly flags this site because the adjudicator grounds its verdict in multiple evidence streams:
(1) \textbf{Visual Understanding:} The operator's VLM visually identifies the semantic presence of a login form and urgency text despite the lack of brand assets.
(2) \textbf{Trace Evidence:} The adjudicator detects obfuscated JavaScript loading from a suspicious \texttt{r2.dev} storage bucket (\emph{T1056.002 - Input Capture: GUI Input Capture}) and notes a lack of legitimate enterprise asset references.

\parhead{Case Study: Abuse of Legitimate Services (Google Forms)}
We also observed attackers hosting scams on legitimate infrastructure, such as Google Forms, to bypass domain reputation filters (Figure~\ref{fig:case-googleform}). One campaign presented a ``2025 BENEFIT PROGRAM'' soliciting Name, Address, and Social Security Number (PII). Because the domain is trusted (\texttt{docs.google.com}) and the page uses standard Google styling without impersonating a third-party brand, snapshot baselines classify it as benign. \system detects the \emph{semantic intent} (soliciting PII for a ``benefit program'') and flags the mismatch between the high-sensitivity request and the unverified hosting context.

\subsubsection{Detecting Non-Standard Objectives (Non-CRP)}
Traditional detectors are optimized for Credential-Requiring Pages (CRP), specifically those soliciting username and password pairs. Consequently, they struggle when the adversary's objective shifts to cryptocurrency authorization or PII harvesting. \system achieved 100\% recall (12/12) on this subset, whereas the best baseline reached only 8.3\%.

\begin{figure}[t]
  \centering
  \includegraphics[width=\linewidth]{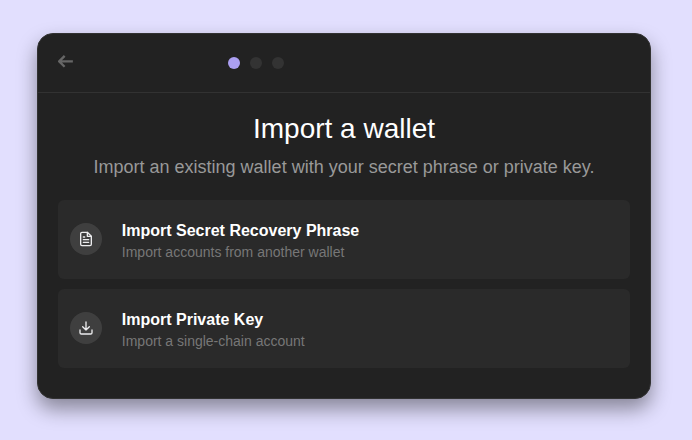}
  \caption{\label{fig:case-crypto}Case study: crypto-wallet phishing without a traditional CRP. The page presents an exchange-like UI and prompts a wallet connection (e.g., MetaMask/Coinbase-style modal) instead of a username/password login form. TraceScope bases its verdict on the interaction-revealed intent, hosting context, and corroborating runtime artifacts.}
\end{figure}

\parhead{Case Study: Crypto-Wallet Drainers}
Figure~\ref{fig:case-crypto} illustrates a ``Wallet Connect'' attack mimicking a cryptocurrency exchange. Rather than presenting a password form, the site displays a modal to link a MetaMask or Coinbase wallet. \system recognizes this intent through interaction. The agent logs the ``Connect Wallet'' prompts, and the adjudicator correlates these events with the hosting domain (\texttt{coinxpd[.]cc}), noting its lack of reputation as a legitimate exchange.

\parhead{Case Study: Conversational Phishing}
Certain attacks employ ``Progressive Disclosure'' techniques that conceal malicious requests behind benign chat interfaces, as shown in Figure~\ref{fig:case-chat}. We observed a campaign mimicking customer support that requested a Brazilian CPF (national ID) only after several turns of scripted dialogue. While static snapshots capture only the initial greeting screen, the \system agent actively engages with the chat. This interaction populates the timeline with the eventual PII request, enabling the adjudicator to analyze the complete attack lifecycle.

\subsubsection{Interaction Gating and Failure Analysis}
\label{sec:interaction_gating}

\begin{figure}[t]
  \centering
  \begin{subfigure}[b]{0.61\linewidth}
    \centering
    \includegraphics[width=\linewidth]{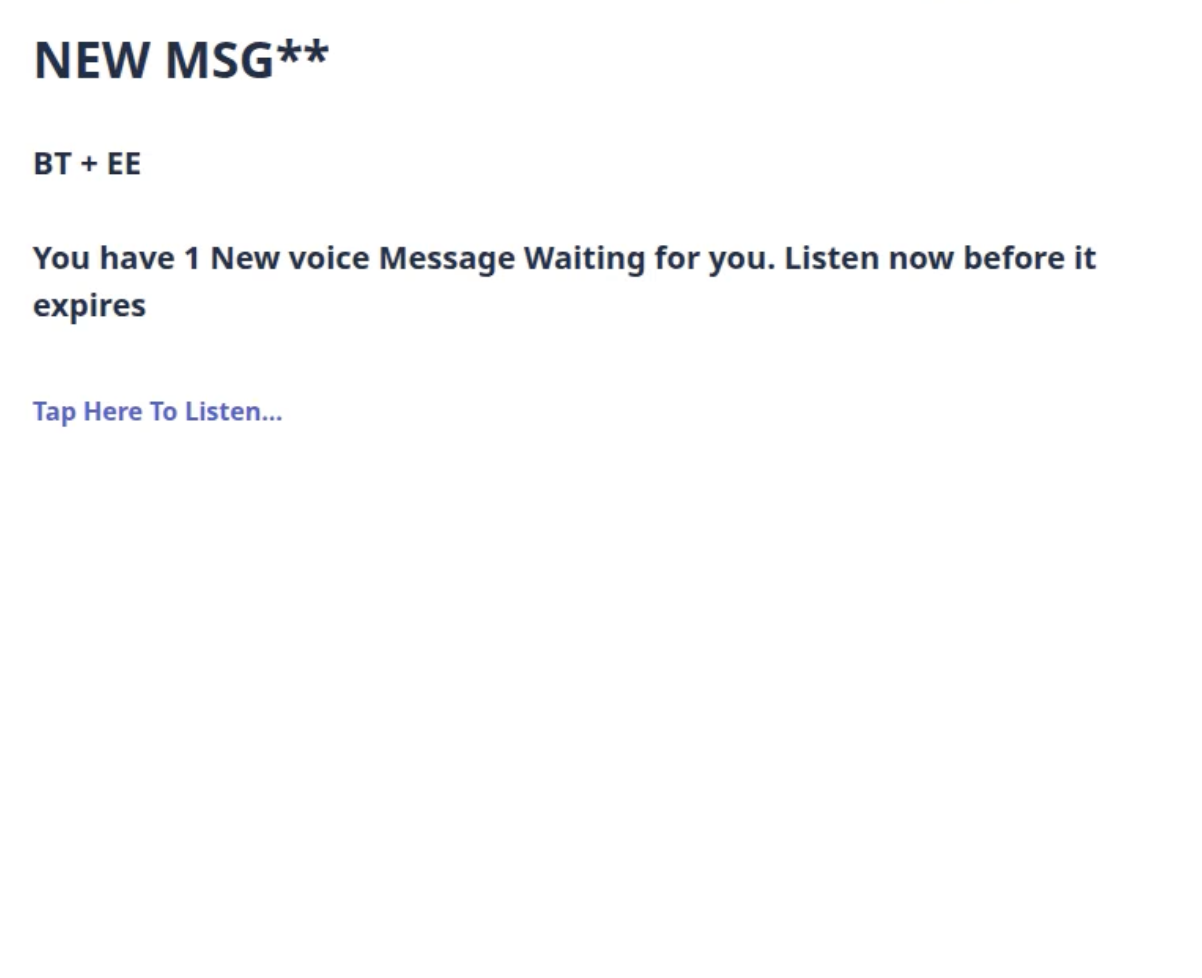}
    \caption{Initial Gate}
    \label{fig:audio_gate}
  \end{subfigure}
  \hfill
  \begin{subfigure}[b]{0.38\linewidth}
    \centering
    \includegraphics[width=\linewidth]{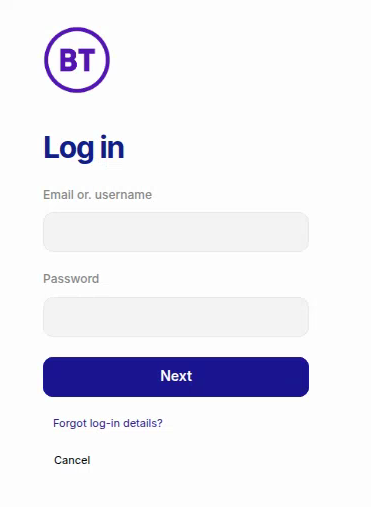}
    \caption{Revealed Payload}
    \label{fig:audio_payload}
  \end{subfigure}
  
  \caption{\label{fig:advanced_interaction} Success Case: Contextual Audio Lure. The phishing content is hidden behind a ``Voice Message'' interaction gate. Static classifiers see only the benign notification (a). \system correctly interprets the ``Tap Here To Listen'' prompt as a required interaction, clicking through to reveal the malicious login payload (b).}
\end{figure}

The ``Interaction Gated'' category (CAPTCHAs, media lures, turnstiles) remains the most challenging for all systems. While \system matched the best baseline ($\text{Recall} \le 16.7\%$), our qualitative analysis reveals distinct boundaries between agent success, instruction-following failures, and adversarial blocking.

\parhead{Success: Decloaking Contextual Lures}
Attackers frequently hide payloads behind benign-looking ``media'' interactions to defeat visual classifiers. We observed campaigns presenting fake voicemail notifications (Figure~\ref{fig:advanced_interaction}) where the malicious URL is only loaded after a specific user action.
In this scenario, static snapshotters see only a benign ``New Message'' prompt. \system successfully identifies the ``Tap Here To Listen'' Call-to-Action, clicks the link, and exposes the underlying credential harvester. This demonstrates the agent's ability to navigate semantic gates that lack standard CAPTCHA labeling.

\parhead{Success: Logic and Arithmetic Gates}
As introduced in Figure~\ref{fig:intro-fig}, \system can traverse logic gates that require external knowledge. When presented with a numeric challenge (e.g., $8+7=$), the agent utilizes its VLM backend to transcribe the prompt, compute the sum ($15$), and simulate the keystrokes to solve the gate.

\begin{figure}[t!]
  \centering
  \begin{subfigure}[b]{0.80\linewidth}
    \centering
    \includegraphics[width=\linewidth]{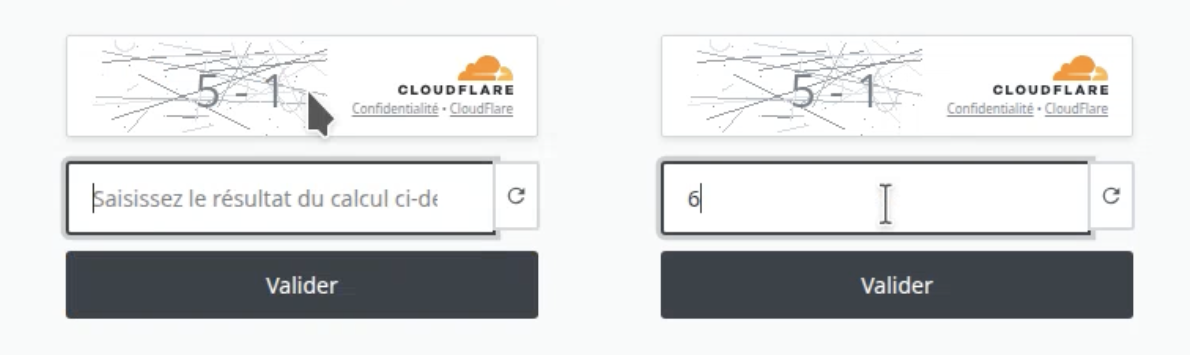}
    \caption{Symbol Hallucination}
    \label{fig:fail_symbol}
  \end{subfigure}
  \hfill
  \begin{subfigure}[b]{0.80\linewidth}
    \centering
    \includegraphics[width=\linewidth]{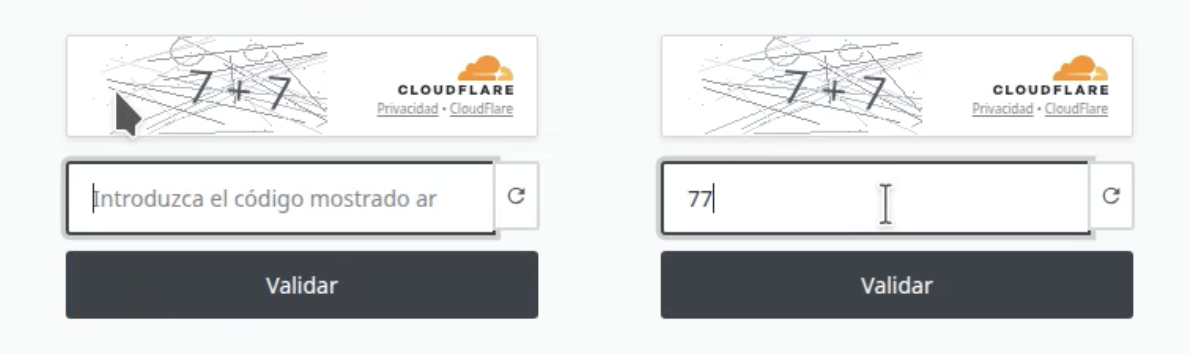}
    \caption{Literal Transcription}
    \label{fig:fail_literal}
  \end{subfigure}
  \caption{\label{fig:captcha_fails} TracePilot failure modes during CAPTCHA solving. In (a), the agent misinterprets the operator, solving $5+1=6$ instead of $5-1=4$. In (b), the agent fails to perform the calculation, literally transcribing the digits ``77'' instead of the sum ``14''.}
\end{figure}

\parhead{Failure Mode: VLM Instruction Following}
Despite these successes, we identified specific failure modes rooted in VLM limitations rather than system architecture (Figure~\ref{fig:captcha_fails}).
First, we observed \emph{Symbol Hallucinations} in low-contrast CAPTCHAs. In Figure~\ref{fig:captcha_fails}(a), the model misrecognized the minus sign ($-$) as a plus ($+$), inputting ``6'' ($5+1$) instead of the correct answer ``4''.
Second, we noted \emph{Literal Transcription Errors} where the agent failed to distinguish the instruction from the input. As shown in Figure~\ref{fig:captcha_fails}(b), given the visual prompt ``$7+7$'', the agent typed the literal digits ``77'' into the input box instead of performing the calculation. These failures suggest that while the \emph{mechanism} for solving exists, robustness depends heavily on the underlying model's ability to adhere to ``solve'' vs. ``transcribe'' instructions.

\parhead{Failure Mode: Adversarial Loops}
Finally, we differentiate agent errors from hard adversarial blocks. As discussed in Section 6 (Figure~\ref{fig:decloak-loop}), some servers enforce infinite challenge loops or return opaque errors even when the agent provides the correct solution. We classify these as ``Blocked'' outcomes, representing a current hard boundary for software-driven analysis distinct from the agent's reasoning capability.

\subsection{RQ3: Operational Efficiency and Viability}
\label{sec:cost_analysis}

\begin{figure}[t]
  \centering
  \includegraphics[width=0.90\linewidth]{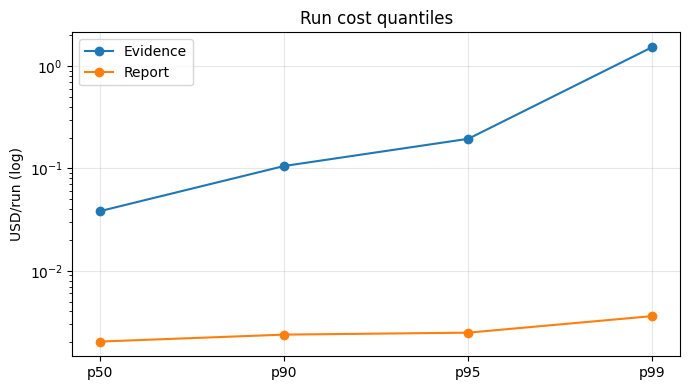}
  \caption{\label{fig:cost_quantiles} Log-scale cost distribution for the TraceSleuth (excluding the constant $\approx$\$0.20 TracePilot cost). The profile is bimodal: a low-cost baseline (\$0.04) for typical sites and dynamic scaling (up to \$1.53) for adversarial targets, confirming adaptive resource allocation.}
\end{figure}

We evaluate the operational overhead based on production telemetry from the OpenRouter platform. In this deployment, the upstream TracePilot utilized Claude-Sonnet-4.5 with a strict 60-second execution cap, yielding a consistent baseline cost of $\approx$\$0.20 per URL. Consequently, we focus our analysis on the downstream TraceSleuth, powered by xAI's Grok-4.1-Fast, where token consumption fluctuates significantly based on the volume of captured evidence. Figure~\ref{fig:cost_quantiles} illustrates this distribution, revealing that resource consumption is highly adaptive to target complexity.

\parhead{Adaptive Resource Allocation}
The system exhibits a bimodal cost profile driven by the site's defensive posture. In the \textit{Standard Regime} (benign or simple phishing), the agent quickly verifies safety, resulting in a median execution cost of \$0.04 per URL. However, in the \textit{Adversarial Regime} (e.g., CAPTCHA loops, obfuscated scripts), the agent automatically escalates its investigative depth. This long-tail behavior (p99 $\approx$ \$1.53) demonstrates that \system dynamically allocates ``cognitive budget'' where it is strictly necessary, avoiding waste on obvious targets.

\parhead{Economic Viability vs. Human Analysts}
While large language model inference is often viewed as costly, \system offers significant savings compared to manual triage. Assuming a human expert analyst wage of \$40/hour, a standard 2-minute manual review costs $\approx$\$1.33. \system's median cost (\$0.24) represents a 5.5$\times$ reduction in operational expense. Even in the most complex p99 cases (\$1.73), the system remains cost-competitive against the 10+ minutes a human would require to de-obfuscate equivalent threats.


\subsection{RQ4: Real World Deployment}
\label{sec:deployment}

\begin{table*}[t]
\centering
\small
\setlength{\tabcolsep}{0pt} 
\caption{\label{tab:deployment_results} Performance during the 14-day real-world pilot. We report metrics on the \textbf{Accessible Subset} (where baselines could render the page) and the \textbf{Real-World Set} (which includes 18 live URLs that blocked baseline crawlers). While baselines collapse when exposed to anti-bot evasion (False Downtime), \system maintains high recall and precision across both scenarios.}
\begin{tabular*}{\textwidth}{@{\extracolsep{\fill}}l ccc c ccc}
\toprule
& \multicolumn{3}{c}{\textbf{Accessible Subset (N=53)}} && \multicolumn{3}{c}{\textbf{Real-World Triage (N=71)}} \\
\cmidrule(r){2-4} \cmidrule(l){6-8}
\textbf{Method} & \textbf{Precision} & \textbf{Recall} & \textbf{F1} && \textbf{Precision} & \textbf{Recall} & \textbf{F1} \\
\midrule
\system & \textbf{1.0000} & \textbf{0.6765} & \textbf{0.8070} && \textbf{0.9375} & \textbf{0.6522} & \textbf{0.7692} \\
PhishIntention~\cite{liu2022phishintention} & 1.0000 & 0.1471 & 0.2564 && 0.4545 & 0.1087 & 0.1754 \\
Phishpedia~\cite{lin2021phishpedia} & 1.0000 & 0.0882 & 0.1622 && 0.3333 & 0.0652 & 0.1091 \\
PhishVLM~\cite{liu2024phishllm} & 0.8000 & 0.1176 & 0.2051 && 0.3636 & 0.0870 & 0.1404 \\
\bottomrule
\end{tabular*}
\end{table*}

To validate \system beyond controlled datasets, we conducted a two-week pilot deployment from January 20th to February 3rd. We collaborated with senior Email Security Analysts who serve as technical leads for IT security compliance and analysis within the organization. In this workflow, analysts manually forwarded suspicious URLs from their daily triage queue to our pipeline, accompanied by brief annotations justifying their suspicion (e.g., ``Bitcoin campaign'', ``fake Apple login page''). This process ensured that the input stream consisted of high-quality, human-verified samples rather than raw, noisy feeds.

\parhead{Dataset Construction and False Downtime}
During the 14-day period, analysts submitted a total of 111 URLs. Consistent with the volatility observed in our live evaluation, a significant portion of these links decayed rapidly; 40 URLs were definitively unreachable (dead DNS or removed by the hosting platform) by the time our system attempted to access them. This left a valid benchmark set of 71 URLs (46 phishing, 25 benign).

However, a naive reachability analysis hides a critical layer of evasion. As noted in recent measurements by Ji et al.~\cite{ji2025_visual_similarity_eval}, a large percentage of URLs that appear dead to crawlers are merely cloaked. We observed this ``false downtime'' explicitly: 18 of the 71 valid URLs (25.3\%) returned HTTP 403/500 errors or connection timeouts when accessed by the headless Selenium drivers used in standard baselines. These sites remained fully functional for human users and our full-stack browser, but successfully fingerprinted and blocked the automated scrapers.

\parhead{Detection Performance}
To rigorously quantify the impact of this evasion, we evaluate performance in two contexts: (1) a \textit{Real-World Triage} setting (N=71), which includes all valid URLs and penalizes systems for failing to access a live threat; and (2) an \textit{Accessible Subset} (N=53), which excludes the 18 contested URLs to provide a theoretically ``fair'' comparison on sites where baselines could successfully render content. Table~\ref{tab:deployment_results} presents the comparative results.

\textit{Scenario A: Accessible Subset.} Even when restricting the evaluation to the 53 URLs accessible to all tools, \system significantly outperforms baselines. While PhishIntention and Phishpedia achieve perfect precision (1.0), their recall is critically low (0.1471 and 0.0882, respectively). This confirms that even without anti-bot cloaking, static visual snapshots fail to capture the multi-step attacks prevalent in this dataset. By interacting with the page, \system achieved a recall of 0.6765, effectively quadrupling the detection rate of the nearest baseline.

\textit{Scenario B: Real-World Triage.} The limitations of current baselines become starkest in the full dataset (N=71). Because the baselines could not render the 18 cloaked websites, they effectively treated them as benign or non-existent, driving their recall down to negligible levels ($\approx$0.10). Furthermore, their precision degraded significantly (e.g., PhishIntention dropped from 1.0 to 0.45), indicating that partial rendering of evasive sites leads to erroneous verdicts. In contrast, \system maintained robust performance (F1 0.7692), validating that a full-stack, interactive browser is a prerequisite for visibility into modern phishing campaigns.

\section{Discussion}

\parhead{Operational Limitations} While \system effectively navigates common interaction gates, we observe specific failure modes driven by aggressive provider-side bot detection. As illustrated in Figure~\ref{fig:decloak-loop}, certain implementations enforce infinite challenge loops where the server presents new gates despite the operator correctly solving the slider or puzzle. This behavior is consistent with prior observations in CAPTCHA-cloaking research~\cite{teoh2024phishdecloaker} and currently represents a hard boundary for software-driven agents. Furthermore, our evaluation results reflect a necessary selection bias, as we strictly exclude unreachable or heavily blocked URLs to ensure dataset consistency. We report the full filtering funnel to provide transparency regarding this constraint.

\parhead{Adversarial Evolution and Future Work} Beyond immediate operational constraints, we acknowledge that the arms race between analysis automation and adversarial evasion is continuous. Current browser automation, including our containerized \texttt{webtop} implementation, leaks subtle artifact signatures. For instance, our telemetry indicates that the sandboxed environment exposes a software-rendering signature (e.g., \texttt{SwiftShader Device}) in the \texttt{WebGL Renderer} string, creating a distinct discrepancy compared to the hardware-accelerated GPUs (e.g., \texttt{Apple M3}, \texttt{NVIDIA}) found in standard consumer devices. Sophisticated adversaries can exploit these signals to cloak malicious payloads and serve benign content to the analysis agent.

To address these vulnerabilities, a primary direction for future work is the transition from containerization to snapshot-based virtualization with \textit{fingerprint normalization}. Future iterations of \system will replace ephemeral containers with persistent QEMU/KVM virtual machines equipped with hardware passthrough. By mimicking specific consumer hardware signatures and ``averaging'' performance metrics to statistically align with the standard user population, we can eliminate the virtualization distinctiveness that modern anti-bot systems target. 
In extreme evasion scenarios where virtualization itself remains detectable, we identify the integration of dedicated hardware farms as the ultimate countermeasure. Orchestrating physical devices via KVM-over-IP interfaces would render the analysis infrastructure indistinguishable from the targeted victim demographic. While operationally expensive, this physical air-gap architecture represents the necessary evolution for high-fidelity threat intelligence.


\section{Conclusion}
TraceScope reframes phishing URL classification as an interactive, evidence-driven triage task. By combining a GUI operator agent with a checklist-scoped adjudicator that reasons over recorded artifacts (and produces audit-ready reports), TraceScope improves recall on live URLs where snapshot classifiers fail due to interaction gates and missing brand cues. Our ongoing work focuses on broader ablations, stronger handling of bot-detection loops, and measuring analyst utility of generated reports.

\newpage
\section*{Ethical Considerations}
In accordance with USENIX Security policies, this appendix provides a stakeholder-based ethics analysis of our research procedures. We identified primary stakeholders as the university users and IT staff who provided the reported url dataset, the research team exposed to malicious content, and the broader security community. While publication provides defenders with tools to counter interactive phishing, we acknowledge that adversaries may adapt their evasion techniques in response. Our decision to publish is grounded in the principle of Beneficence, as the community lacks robust defenses against cloaked phishing attacks that currently bypass standard filters.

To respect the privacy of individuals, all emails and URLs were de-identified by removing names, department headers, and unique tracking tokens before research access. This retrospective evaluation involved no human subject contact or active deception. Furthermore, all browser interactions by TracePilot occurred within ephemeral, air-gapped containers to mitigate malware risks. The use of synthetic personas ensured that no real user credentials or financial data were at risk of being harvested. Finally, our browser interactions were limited to eliciting behavior from URLs already flagged by users, ensuring respect for the law and public interest while prioritizing the safety of the internet ecosystem.

\section*{Open Science}
To comply with the USENIX Security Open Science Policy and to support the reproducibility of our findings, this appendix details the artifacts provided to evaluate the core contributions of this paper. We have made the following artifacts available:

\begin{itemize}
    \item Code: Source code for the proposed pipeline framework.
\end{itemize}

The dataset used in this study is not included in the artifact submission, as the core contribution of this work lies in the pipeline implementation rather than the data itself.

In accordance with the double-blind review process, the provided artifacts are hosted on an anonymized repository and can be accessed via:

\begin{center}
    \url{https://anonymous.4open.science/r/tracescope-AF2C}
\end{center}

To execute the code and reproduce the results, valid API keys for the cloud-hosted LLM vendors are required. No local LLM environment is necessary for this implementation.

\newpage
\bibliographystyle{plain}
\bibliography{ref}

\newpage
\appendix

\section{System Prompts and Agent Instructions}
\label{app:prompts}

This appendix provides the system prompts used to initialize the core agents within the \system architecture. These instructions are critical for reproducibility, as they define the behavioral guardrails, output structures, and persona constraints that govern the autonomous workflow.

\begin{figure}[ht!]
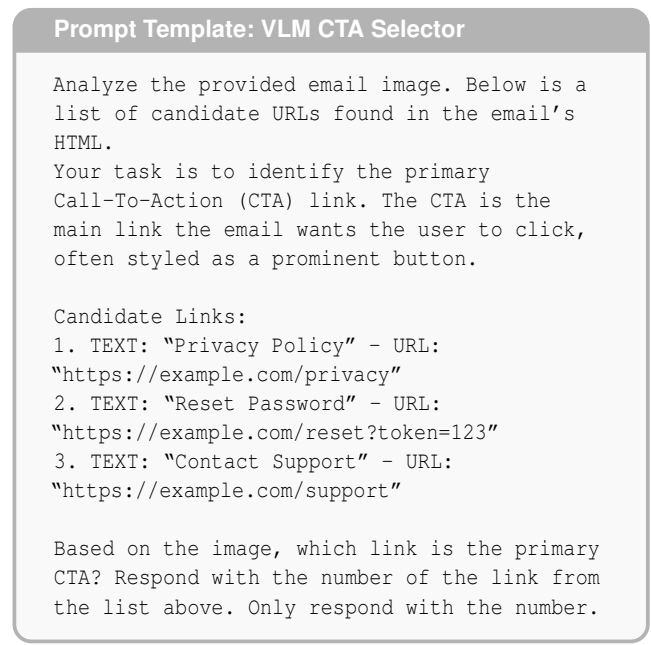

    \centering
    \begin{tcolorbox}[
        colback=gray!5,       
        colframe=gray!60,     
        title=\textbf{Prompt Template: VLM CTA Selector}, 
        fonttitle=\bfseries\sffamily,
        boxrule=0.5mm,
        arc=2mm               
    ]
    \small\ttfamily           
    
    Analyze the provided email image. Below is a list of candidate URLs found in the email's HTML.\\
    Your task is to identify the primary Call-To-Action (CTA) link. The CTA is the main link the email wants the user to click, often styled as a prominent button.\\
    
    Candidate Links:\\
    1. TEXT: ``Privacy Policy'' - URL: ``https://example.com/privacy''\\
    2. TEXT: ``Reset Password'' - URL: ``https://example.com/reset?token=123''\\
    3. TEXT: ``Contact Support'' - URL: ``https://example.com/support''\\
    
    Based on the image, which link is the primary CTA?
    Respond with the number of the link from the list above. Only respond with the number.
    \end{tcolorbox}
    
    \caption{The zero-shot prompt template used to guide the Vision-Language Model. The model receives this text prompt alongside the sanitized email screenshot.}
    \label{fig:vlm_prompt}
\end{figure}

Figure~\ref{fig:vlm_prompt} details the zero-shot prompt used by the Email Preprocessor to semantically identify the Call-to-Action (CTA) link, while Figure~\ref{fig:report_prompt} defines the structure for the final human-readable incident report.

\begin{figure}[ht!]
    \centering
    \begin{tcolorbox}[
        colback=green!5,      
        colframe=green!60,    
        title=\textbf{System Prompt: Incident Report Writer}, 
        fonttitle=\bfseries\sffamily,
        boxrule=0.5mm,
        arc=2mm
    ]
    \small\ttfamily
    You are an expert Incident Response Report Writer. Your goal is to turn structured findings into a polished, actionable report.\\[1em]
    
    Critical Style Guidelines:\\
    - \textbf{NO REPETITION}: Do not repeat facts across sections.\\
    - \textbf{Action-Oriented}: Focus on ``what happened'' and ``what to do next.''\\
    - \textbf{Evidence-Based}: Cite tool invocation IDs explicitly.\\[1em]
    
    Report Template Structure:\\
    1. \textbf{Executive Summary}: Verdict (Phishing/Safe), Confidence, Rationale.\\
    2. \textbf{Scope \& Context}: Assets and accounts involved.\\
    3. \textbf{Timeline (UTC)}: Ordered list of key events with timestamps.\\
    4. \textbf{IOCs}: Domains, IPs, hashes supported by data.\\
    5. \textbf{Risk Assessment}: Impact and blast radius.\\
    6. \textbf{Actionable Recommendations}: Prioritized ``To-Do'' checklist.\\
    7. \textbf{Appendices}: Evidence Cross-Reference, MITRE ATT\&CK Mapping.
    \end{tcolorbox}
    
    \caption{The prompt configuration for the final reporting phase. The agent converts raw JSON verdicts into a standardized Incident Response document, prioritizing executive clarity and actionable remediation steps over technical verbosity.}
    \label{fig:report_prompt}
\end{figure}

\begin{figure}[t]
  \centering
  \includegraphics[width=\linewidth]{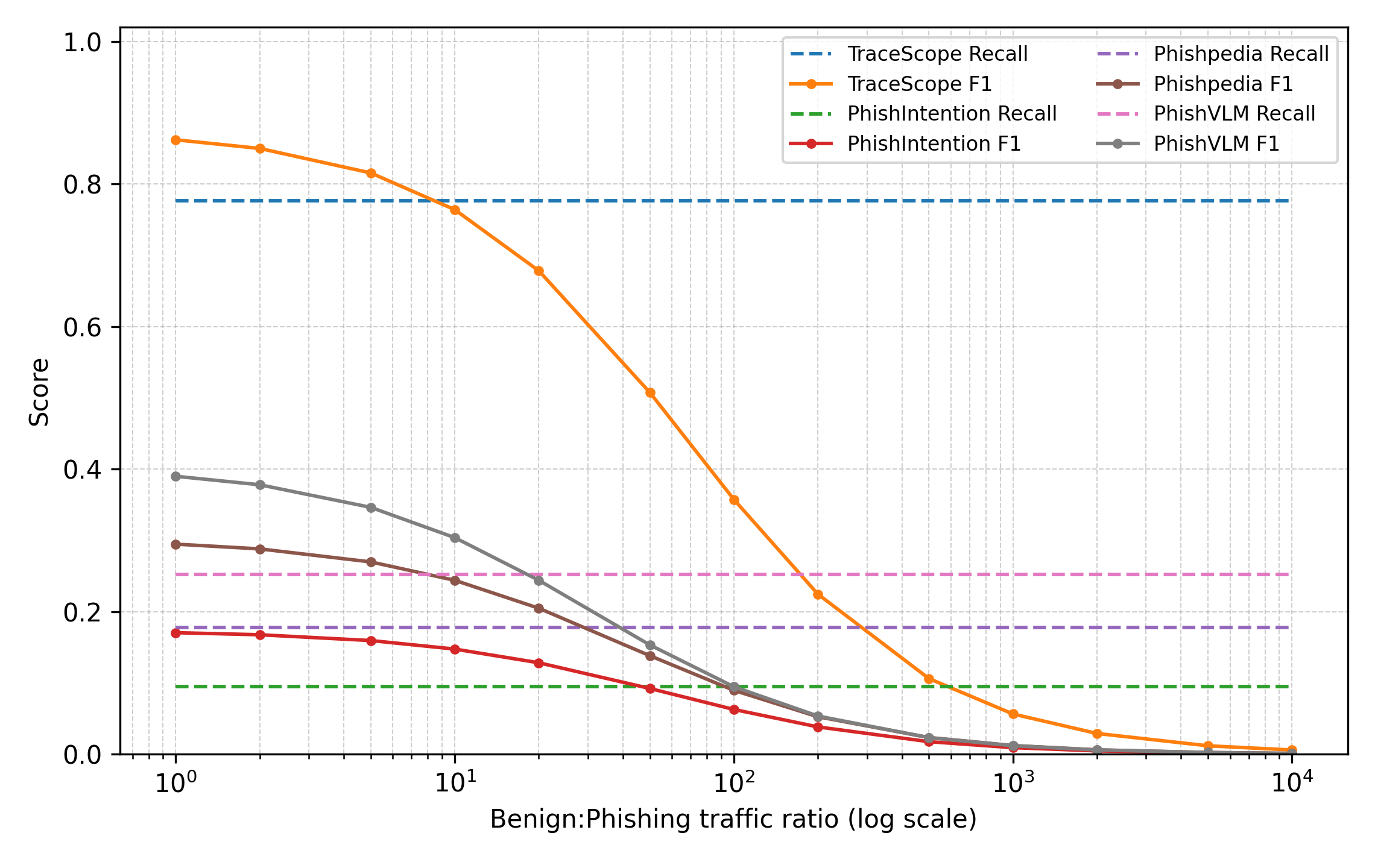}
  \caption{\label{fig:prior_shift}
  projected F1-score under class-prior shift. \system maintains higher utility across all prevalence ratios, though it is optimized for the high-prevalence setting of an analyst triage queue.}
\end{figure}

\section{Example of Summarized Report}
\label{app:sample_report}

\begin{figure*}[t]
  \centering
  \includegraphics[width=0.85\textwidth]{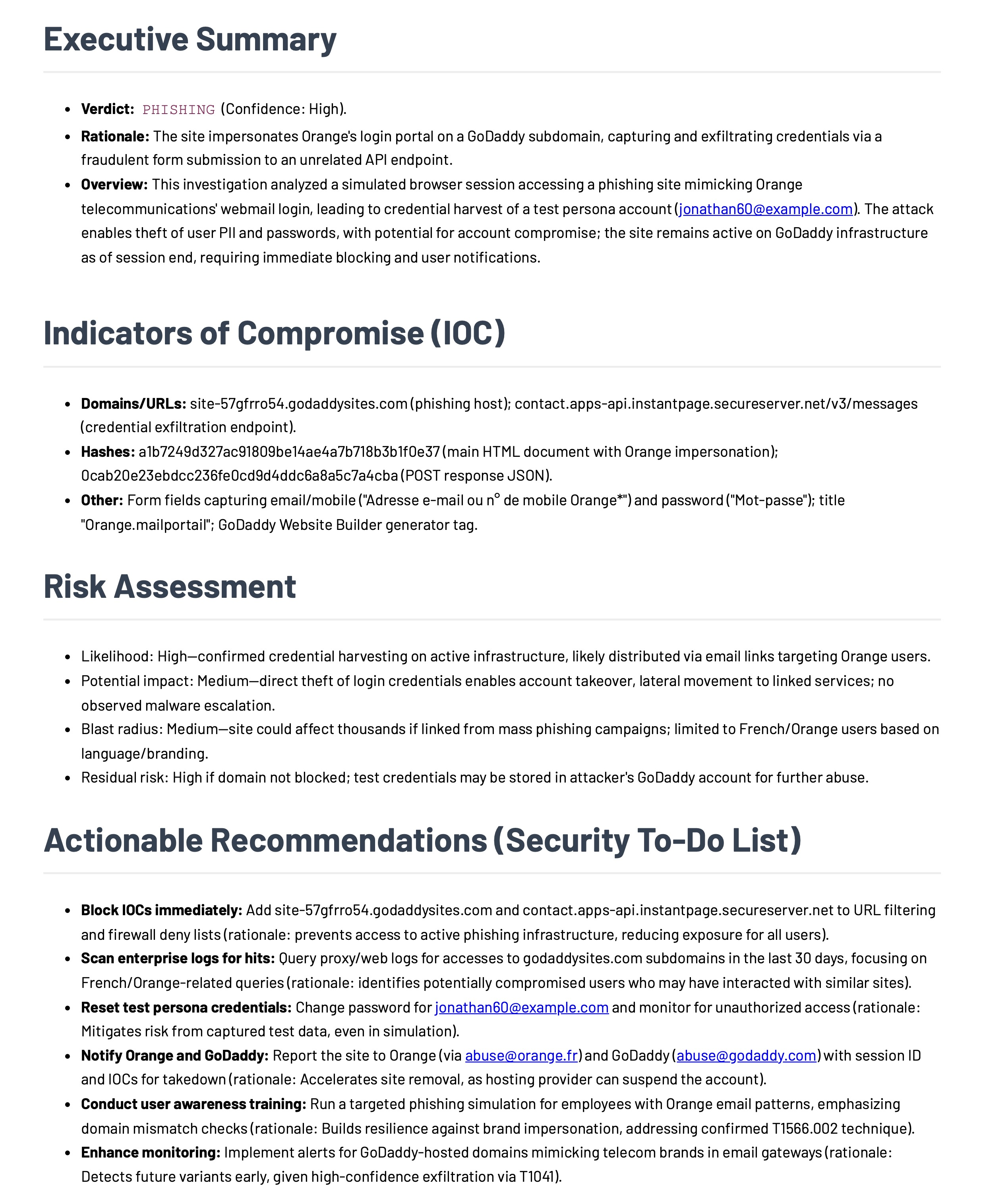}
  \caption{\label{fig:report} A condensed view of the automated incident report for the GoDaddy-hosted phishing campaign. The system synthesizes the adjudication verdict into a standardized BLUF format, automatically extracting high-priority IOCs (e.g., the exfiltration endpoint) and generating actionable containment steps for security analysts. \textit{Note: The detailed evidence timeline, methodology, and raw artifact appendices have been omitted from this figure for brevity.}}
\end{figure*}

To demonstrate the system's end-to-end utility, Figure~\ref{fig:report} displays a condensed representative incident report generated by \system. This output translates the raw JSON findings from the adjudication phase into a standardized, human-readable document. The report prioritizes ``Bottom Line Up Front'' (BLUF) information for executive decision-making, followed by technical indicators of compromise (IOCs) and actionable remediation steps. By synthesizing complex forensic data into immediate strategic guidance, the system allows analysts to validate the verdict and initiate containment without manually parsing raw session logs.


\section{Detailed Analysis of Failure Modes}
\label{app:failure_modes}

\begin{figure*}[t]
  \centering
  \includegraphics[width=\linewidth]{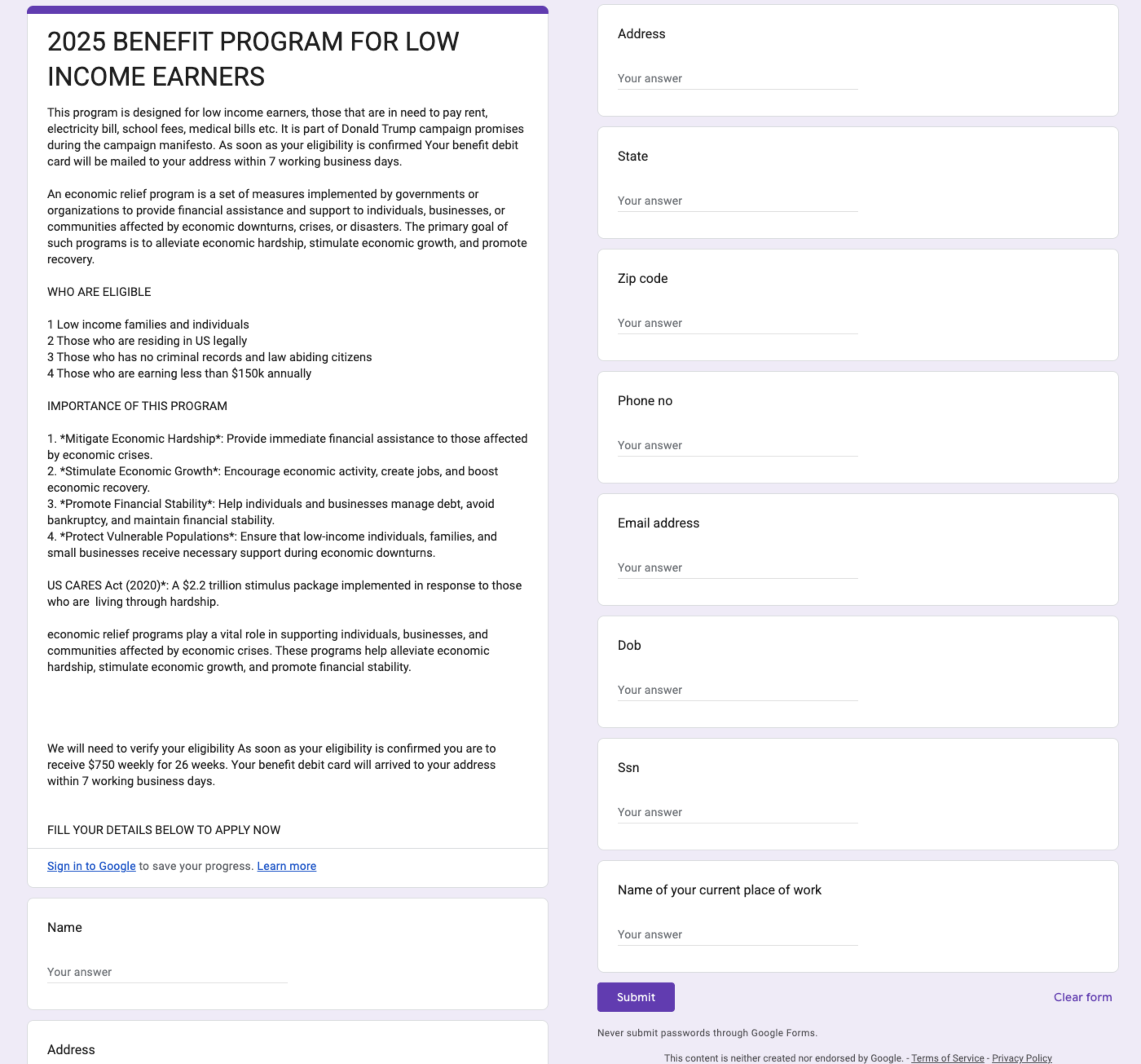}
  \caption{\label{fig:case-googleform}Case study: PII-harvesting phishing abusing Google Forms. The attacker hosts scam content on a legitimate Google Forms URL and solicits PII (e.g., name and address), which can evade logo-centric and domain-centric detectors.}
\end{figure*}

\begin{figure*}[t]
  \centering
  \includegraphics[width=\linewidth]{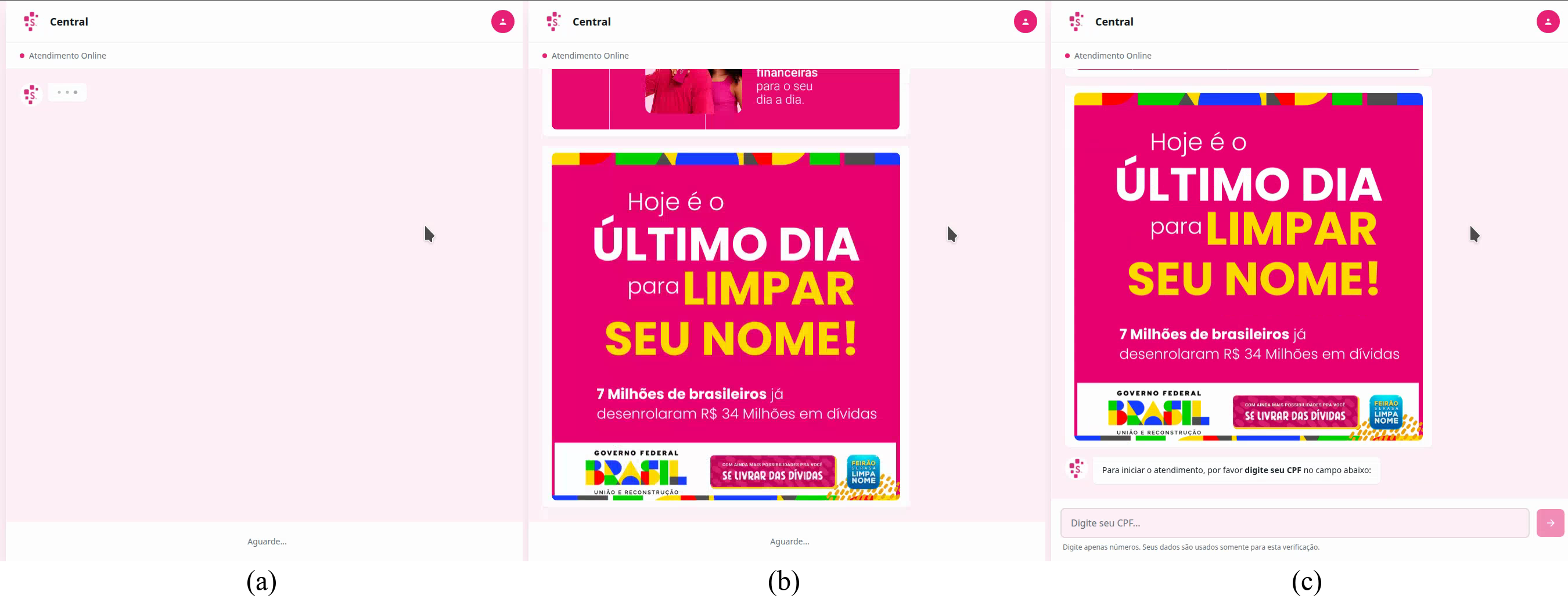}
  \caption{\label{fig:case-chat}{Case study: conversational ``support chat'' phishing via progressive disclosure. (a) Initial landing state shows an ``online service'' chat interface with a typing indicator and no explicit credential form. (b) The page dynamically renders scripted prompts to establish legitimacy. (c) After several turns, the chat requests Brazilian CPF (a national identifier) as sensitive PII, despite no clear relationship to a trusted domain or verified service. TraceScope captures the interaction timeline and corroborates intent using runtime evidence (rendered states, loaded resources, and submission destinations), enabling a phishing verdict where CRP-focused baselines under-trigger.}
}
\end{figure*}

This section provides visual case studies of the complex evasion techniques and failure modes discussed in Section~\ref{sec:rq2} and Section~\ref{sec:interaction_gating}.

\begin{figure*}[t]
  \centering
  \includegraphics[width=\textwidth]{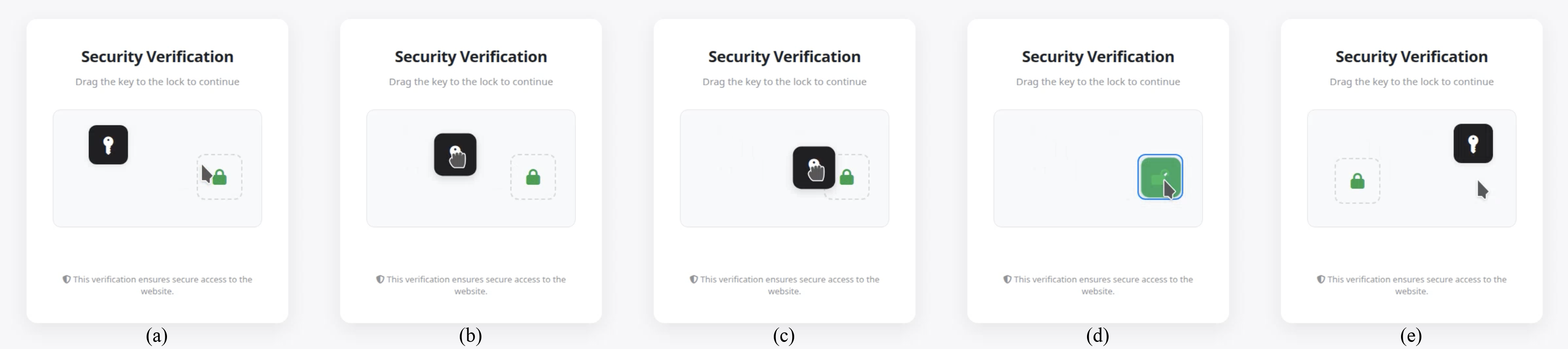}
  \caption{\label{fig:decloak-loop}Challenge-loop failure mode during interactive decloaking. Panels (a--d) show the operator agent correctly completing a slider-style challenge (dragging the key to the lock). In (e), the server presents another challenge instead of revealing downstream content, resulting in an infinite loop consistent with provider-side bot detection.}
\end{figure*}

\begin{figure*}[t]
  \centering
  \includegraphics[width=\linewidth]{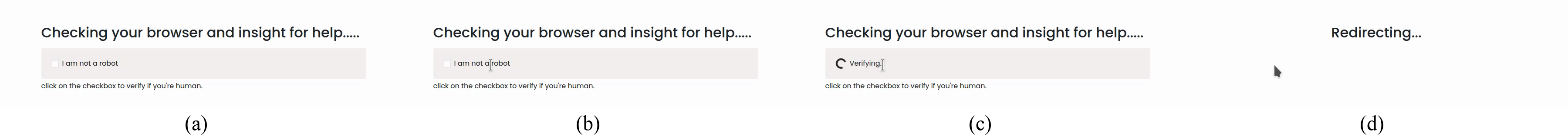}
  \caption{\label{fig:case-captcha}CAPTCHA-gated landing flow observed during live triage. Panels (a--c) show a checkbox-style challenge and verification state that blocks snapshot crawlers; (d) shows the page transitioning (redirecting) after successful completion, enabling downstream content to load for further analysis.}
\end{figure*}

Figure~\ref{fig:decloak-loop} illustrates an ``infinite loop'' defense, where the adversary repeatedly serves new challenges to exhaust the analysis budget. Figure~\ref{fig:case-googleform} and Figure~\ref{fig:case-chat} highlight the necessity of semantic reasoning: in both cases, visual appearance alone is insufficient to identify the threat. Figure~\ref{fig:case-googleform} abuses a legitimate domain (Google Forms) to bypass reputation filters, requiring the agent to detect the PII-harvesting intent. Figure~\ref{fig:case-chat} uses a multi-turn chat interface to hide the malicious request, which \system exposes through active interaction. Finally, Figure~\ref{fig:case-captcha} demonstrates a successful traversal of a standard checkbox CAPTCHA, a capability that allows \system to analyze content that remains invisible to static crawlers.

\section{Performance Under Distribution Shift}
\label{app:additional_results}

The prevalence of phishing URLs varies significantly depending on the deployment point (e.g., a high-volume perimeter firewall vs. a curated abuse inbox). To characterize \system's stability across these environments, Figure~\ref{fig:prior_shift} models the F1-score as a function of the benign-to-phishing ratio. While all systems show degradation in extremely low-prevalence settings, \system maintains superior utility compared to baselines, particularly in the high-prevalence regimes characteristic of the targeted triage queues for which it is designed.

\end{document}